# Habilitation à Diriger des Recherches
de l'Université Pierre et Marie Curie – Paris VI

## From Spin Torque Nano-Oscillators to Memristors: Multi-Functional Nanodevices for Advanced Computing

2013/06/26

Julie Grollier
Unité Mixte de Physique CNRS/Thales
Palaiseau, France

**Jury composition:**
M. Alain Cappy (IRCICA, referee)
M. Mark Stiles (NIST Gaithersburg, referee)
M. Michel Viret (CEA SPEC, referee)
M. Vincent Cros (UMPhy CNRS/Thales)
M. Giancarlo Faini (LPN-CNRS)
M. Albert Fert (UMPhy CNRS/Thales)
Mme Catherine Gourdon (INSP)
M. Andrei Slavin (Oakland University)

# TABLE OF CONTENTS

# A. Summary / Foreword

In 1996, John Slonczewski and Luc Berger predicted the possibility to manipulate the magnetization of nano-objects by direct current injection, without the need of any applied magnetic field [1] [2]. This phenomena, called spin transfer torque, hold the promise of very interesting research on the fine understanding of the interaction between the spins carried by conduction electrons and local magnetizations. Of course, important applications to non-volatile binary memories were also at stake, since the first generation of Magnetic Random Access Memories (MRAM), in which the magnetization state was written by application of a field, was not technologically viable.

During my Master thesis at UMPhy CNRS/Thales in 2000, we studied spin torque in magnetic nanopillars, and were among the first to demonstrate current-induced magnetization switching at zero field. We also followed through Luc Berger's idea that spin currents could displace magnetic domain walls, and provided a first clear evidence of that effect during my Ph.D. thesis. By digging in the results of spin torque induced magnetization reversal in nanopillars, it soon became apparent that spin torque was not only interesting for its transport properties, but that it could also generate original dynamic magnetization states, in particular sustained magnetic precessions.

The study of the interplay between transport and spin torque induced magnetization dynamics kept me occupied from that point. These spin torque nano-oscillators have a great potential for applications as tiny microwave sources, but their power and spectral purity were not good enough when we started to measure them. My guideline for research since my recruitment in CNRS in 2006 has then been to develop fundamental studies in order to propose practical solutions to these locks. In particular, I suggested synchronizing arrays of spin-torque oscillators connected in series as a solution to both issues. It turned out easier to propose than to achieve, which led us to develop experiments and models of spin torque oscillators phase locking dynamics, then to focus on vortex oscillators. As I will show, the latter are good candidates to finally achieve the synchronization of more than five oscillators.

After all these years working on spin transfer torque I am still amazed by the versatility of this physical effect that allows, by tuning different parameters such as input current waveform, materials (spin valve or magnetic tunnel junction, in-plane or out-of-plane magnetized layers etc.), geometry (pillar, point contact, stripe), to design specific functions at the nano-scale. In 2009, I came to read about the recent results by Hewlett-Packard on memristor nano-devices and their potential applications as nano-synapses in on-chip neural networks. As these perspectives were particularly fascinating I started to think of the possibilities of spin torque in this field, and came up with the idea of a spin torque memristor based on current induced manipulation of a domain wall. After discussions with colleagues working on ferroelectric tunnel junctions showing orders of magnitude resistance variations when flipping the electric polarization, we also started to develop a ferroelectric memristor.

Thanks to this research on memristor nano-devices, I had the chance to start discussing with scientists from other disciplines also extremely interested by the potential of nano-devices for the future developments of hardware neuromorphic architectures. These interactions with enthusiastic computer scientists and architects, neuroscientists, mathematicians, electronic designers and more gave me the virus of inter-disciplinary research. I wish to participate to the extraordinary current research momentum on the understanding, modeling and imitation of the brain. That's why my research project is focused on the development of smart spintronic and non-spintronic nano-devices as building blocks for low power, high performance bio-inspired hardware architectures.

# B. RÉSUMÉ / PRÉAMBULE

En 1996, John Slonczewski and Luc Berger ont prédit qu'il était possible de manipuler l'aimantation de nano-objets par injection de courant à travers la structure, en l'absence de champ magnétique extérieur [1] [2]. Ce phénomène, appelé transfert de spin, ouvrait la voie à de très intéressantes recherches sur les interactions entre les spins portés par les électrons de conduction et les aimantations locales. Bien entendu, d'importantes applications aux mémoires binaires non-volatiles étaient également en jeu, la première génération de mémoires magnétiques MRAM (Magnetic Random Access Memories) utilisant un champ magnétique pour écrire la configuration magnétique n'étant pas technologiquement viable.

Pendant mon stage de Master à l'UMPhy CNRS/Thales en 2000, nous avons étudié le transfert de spin dans des nano-piliers magnétiques, et nous avons été parmi les premiers à démontrer la commutation magnétique sous courant à champ nul. Nous avons également suivi la proposition de Luc Berger d'un déplacement de paroi magnétique sous courant, et nous en avons démontré clairement la possibilité pendant ma thèse. En approfondissant les résultats de renversement d'aimantation par transfert de spin dans les nano-piliers, il est vite apparu que le transfert de spin n'était pas seulement intéressant du point de vue du transport, mais qu'il pouvait également permettre de générer des états dynamiques d'aimantations originaux, en particulier une précession entretenue d'aimantation.

L'étude des interactions entre les propriétés de transport et la dynamique d'aimantation induite par transfert de spin a dès lors constitué mon activité de recherche principale. Ces nano-oscillateurs à transfert de spin ont un fort potentiel applicatif pour une utilisation en tant que nano-source hyperfréquence, mais leur puissance émise ainsi que leur largeur de raie n'étaient pas satisfaisantes lorsque nous avons commencé à les mesurer. Ma ligne de recherche depuis mon recrutement au CNRS en 2006 a été de développer des études fondamentales pour proposer des solutions à ces verrous technologiques. J'ai en particulier suggéré la synchronisation de réseaux d'oscillateurs à transfert de spin connectés électriquement comme solution commune aux deux problèmes. En pratique cette synchronisation s'est révélée plus complexe qu'il n'y paraissait à première vue. Ceci nous a amené à développer des expériences et des modèles de la dynamique du verrouillage de phase d'oscillateurs à transfert de spin, puis à nous intéresser aux girations de vortex magnétique. Comme je le montrerai, ces derniers sont en effet d'excellents candidats pour parvenir enfin à la synchronisation de plus de 5 oscillateurs.

Après toutes ces années de travail sur le transfert de spin, je reste émerveillée par la versatilité de cet effet physique qui permet, en jouant sur des paramètres tels que la forme d'onde du courant injecté, les matériaux utilisés (vanne de spin ou jonction tunnel magnétique, matériaux aimantés dans le plan ou hors du plan etc.), la géométrie (pilier, contact ponctuel, barreau), de réaliser à l'échelle nanométrique des fonctions bien spécifiques. En 2009, j'ai eu connaissance des travaux de Hewlett-Packard sur les nano-composants memristors et leurs applications potentielles en tant que nano-synapse dans des réseaux de neurones artificiels sur silicium. Ces perspectives m'apparaissant particulièrement fascinantes, j'ai commencé à réfléchir aux possibilités du transfert de spin dans ce domaine, ce qui m'a amené à proposer un concept de memristor basé sur la manipulation d'une paroi magnétique par transfert de spin. Par la suite, en discutant avec mes collègues travaillant sur les jonctions tunnel ferroélectriques dont la résistance varie de plusieurs ordres de grandeur lorsque la polarisation est renversée, nous avons eu l'idée du memristor ferroélectrique, que nous avons commencé à développer.

Grâce à ces recherches sur les memristors, j'ai eu la chance de pouvoir échanger avec des scientifiques d'autres disciplines extrêmement intéressés par le fort potentiel des nano-composants pour le futur développement d'architectures neuromorphiques sur puce. Ces interactions avec d'enthousiastes informaticiens, neurobiologistes, mathématiciens, électroniciens et d'autres encore m'ont inoculé le virus de l'interdisciplinarité. Je souhaite participer à l'extraordinaire élan de recherche actuel sur la compréhension, la modélisation et l'imitation du cerveau. Pour ces raisons, mon projet de recherche est orienté sur le développement de nano-composants intelligents comme briques de base d'architectures sur silicium bio-inspirées à faible consommation et haute performance.

# C. Research activities

## I. Spin Transfer Nano-Oscillators

### 1. Introduction

In 1988, Giant Magneto-Resistance (GMR) [3] [4] set the ground to a whole new field in physics: spintronics, and revolutionized data storage through the development of the GMR hard drive read head allowing immense storage capacities (1.8 zetabytes of data stored in 2012 only). But GMR, and more recently Tunnel Magneto-Resistance [5] (TMR) based sensors, are passive elements dedicated to reading-out magnetic states in nano-structures. A mean to actively manipulate the magnetization of nano-objects was provided by spin torque (ST), upgrading spintronic devices to the status of active elements. Indeed, ST, predicted in 1996 [1] [2] and first observed around 2000 [6] [7] [8] [9] allows manipulating efficiently magnetic configurations without the assistance of external magnetic fields (hardly compatible with downscaling) through a simple transfusion of angular momentum from spin-polarized carriers to local magnetic moments. A whole class of new devices, based on the combined effects of ST for writing and GMR or TMR for reading has emerged. The second generation of MRAMs (Magnetic Random Access Memory) based on spin torque writing, called ST-MRAM, is under industrial development and has the potential to replace current cache memories technologies in the next years thanks to its low power consumption and endurance [10]. But spin torque devices are far from being limited to binary memories. As we will see in the following, spin torque allows implementing a variety of functionalities at the nanoscale. In particular spin torque microwave devices seem in good position to be the next candidates on the road to applications. In this section, I will particularly focus on tiny microwave sources: spin torque nano-oscillators.

#### a. Spin torque basics

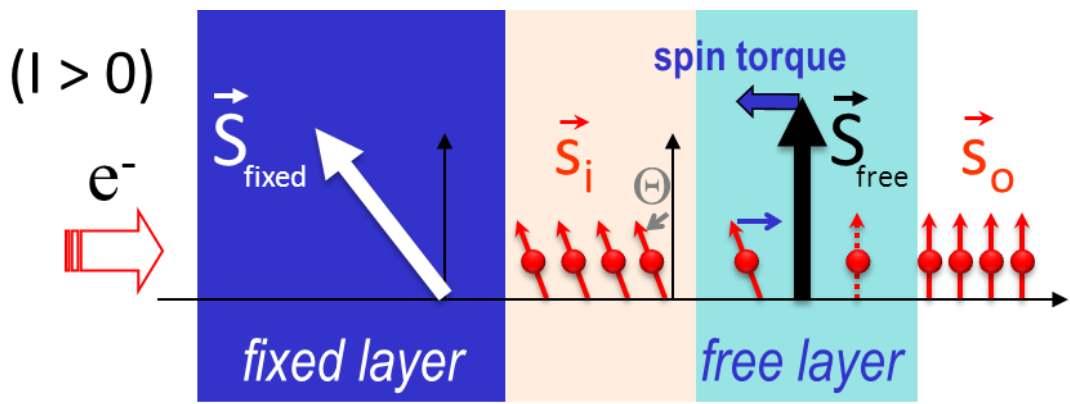

**Figure 1**: Spin torque principle: in a ferromagnet/non-magnetic/ferromagnet trilayer, the transverse spin component of the conduction electrons is absorbed as they pass through the free layer, generating a torque on the local magnetization: the spin-transfer torque.

The archetypal structure of Spin Torque devices is a non-magnetic layer sandwiched between two thin nano-magnets (see Fig. 1). One of the magnetization ($M_{fixed}$) is usually kept fixed, whereas the second one ($M_{free}$) is free to move. It is known from a long time that carriers get spin polarized when passing through ferromagnets [11]. In the case where the magnetizations $M_{fixed}$ and $M_{free}$ are not collinear, the polarized spins incoming on the free layer are not aligned with $M_{free}$. However while passing through the free layer, these spins are going to align with $M_{free}$ due to exchange interaction. During this process at the origin of Spin Torque, the conduction electrons spins have lost their component transverse to $M_{free}$. Given the angular momentum conservation, this lost spin component is transferred to the free layer in the form of a torque: the spin transfer torque, or spin torque. The spin torque can rotate the magnetization of the free layer towards or away from the fixed layer, depending on the sign of the injected current. As predicted by the pioneer works of John Slonczewski [1] and Luc Berger [2], the spin torque amplitude is proportional to the current density, requiring approximately $10^7$ A.cm$^{-2}$ to switch a magnetization at zero field. It is a decisive advantage of spin torque that the smallest the device dimensions are, the lowest is the current needed to manipulate the magnetic state. After a decade of intense research and development, the excellent scalability of Spin Torque has been recently highlighted with low current (< 30 µA) spin torque magnetization switching at room temperature in 20 nm diameter junctions [12].

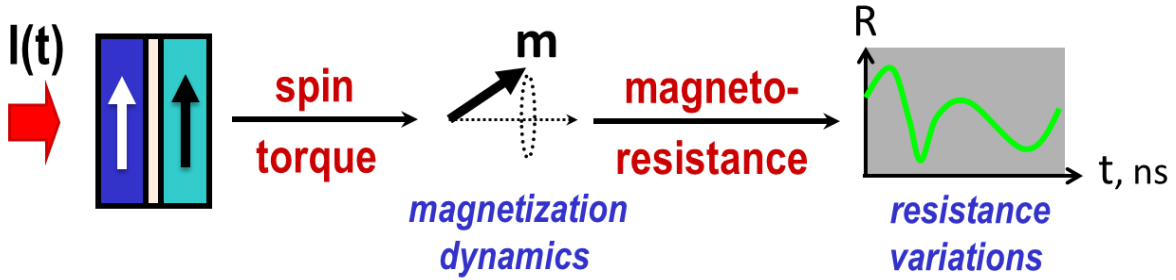

**Figure 2:** General concept of a spin-torque nano-device. As current flows through the trilayer, spin torque induces magnetization dynamics, converted to resistance variations by magneto-resistive effects.

The general principle of spin torque nano-devices is depicted in Fig. 2. (i) A current is injected through the trilayer structure. (ii) Under the action of spin torque, magnetization dynamics can be generated. (iii) This magnetization motion is converted into resistance and voltage variations thanks to the trilayer magnetoresistance, GMR or TMR, depending on the stack.

### b. Spin torques: microscopic origins and consequences

The spin torque has two contributions, called in-plane and out-of-plane torques [13] [14] that provide two different handles to manipulate the magnetization. The in-plane torque $T_{IP}$ lies in the plane defined by $M_{free}$ and $M_{fixed}$, while the out-of-plane torque $T_{OOP}$ points out of it. Their respective origins are illustrated on Fig. 3a and b. Let's consider that a spin polarized current with spins tilted with an angle θ from the z direction, is propagating through the free ferromagnetic layer, where the local spin $S_{free}$ is aligned with z. We examine the origins of the loss of the transverse component of this spin current. As illustrated on Fig. 3a, due to the large exchange field (≈ 1000 T), the spins $s$ carried by the conduction electrons start to precess around $S$ while they propagate through the ferromagnet. Because incident wave vectors have different directions, after a few rotations the coherence is lost and, on average, the transverse spin component is lost. This decoherence process, that gives rise to spin torque, occurs on a length scale $\lambda_\perp$ of the order of 1–2 nm for 3d transition metals like Co [15] [16] [17]. If the free layer thickness t is large compared to $\lambda_\perp$, the outgoing spins $s_o$ are in average ($S_o$) aligned with **S**. The in-plane torque that results from this imbalance can be written as $T_{IP} = -\Delta S = S_i - S_o$.

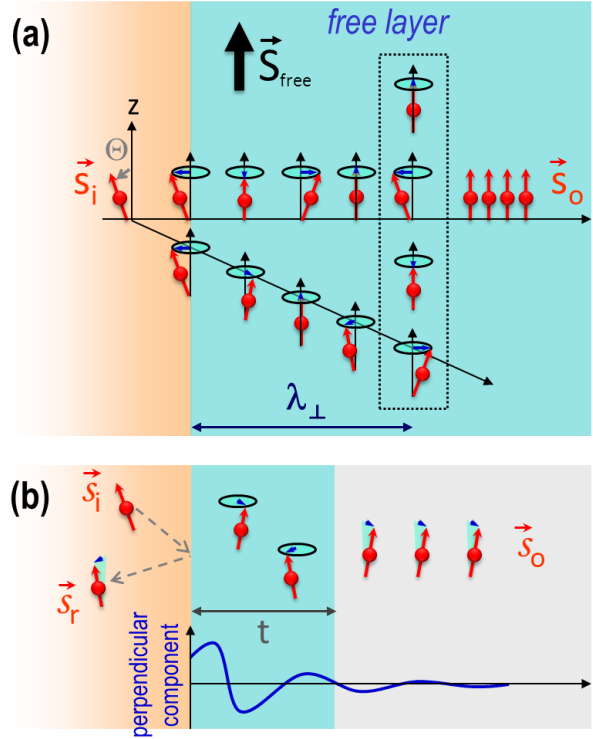

**Figure 3**: Microscopic origins of spin transfer torques (a) In-plane torque: loss of the transverse component of a spin current flowing through a metallic ferromagnet by the combined effects of precession and dephasing. (b) The two contributions to the out-of-plane torque: 1 - the remaining out-of-plane component of transmitted spins if the free layer thickness is smaller than the dephasing length, 2 – the out-of-plane component of reflected spins

The exact expression for the torque spin imbalance equation depends on incoming, outgoing and reflected spins (averaged over the Fermi sphere): $T_{tot} = -\Delta S = \langle s_i + s_r - s_o \rangle$. There are then two possible contributions to the out-of-plane torque, illustrated on Fig. 3b. The first contribution to $T_{OOP}$ is large if the free layer thickness t is smaller than the decoherence length $\lambda_\perp$. In that case the outcoming spins $s_o$ have an out-of-plane component and so does $T_{OOP}$. In magnetic tunnel junctions, due to the filtering of incident k vectors, $\lambda_\perp$ is predicted to be much larger than in SVs, and therefore in most experimental situations, $\lambda_\perp$ is not negligible compared to the free layer thickness t. The second contribution to the out-of-plane torque comes from reflected spins. Due to their interaction with the local spin **S**, reflected spins slightly precess and gain an out of plane component. If the reflected spins $s_r$ conserve some coherence, as it is the case in MTJs, the torque then acquires an out of plane component [14] [18]. That's why, while $T_{OOP}$ is practically zero in metallic spin-valves, it can reach 40 % of $T_{IP}$ in Magnetic Tunnel Junctions [19] [20] [21]. Its current dependence is more complex than $T_{IP}$, as we will see later on.

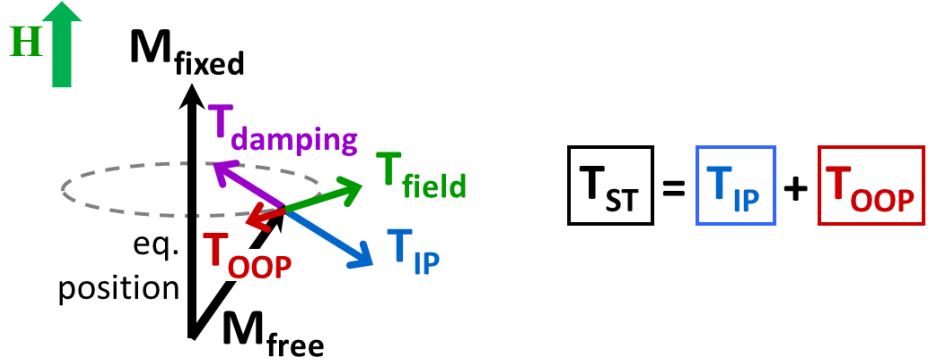

**Figure 4:** Torques on the local magnetization, under current injection, in the particular case where $M_{fixed}$ and the effective magnetic field are aligned. The conservative torques, the out-of plane torque $T_{OOP}$ and the effective-field torque $T_{field}$, and the dissipative torques $T_{IP}$ and $T_{damping}$, are respectively aligned.

The fact that the in-plane torque $T_{IP}$ lies in the plane defined by $M_{free}$ and $M_{fixed}$, while the out-of-plane torque $T_{OOP}$ points out of it, implies very different actions of each torque on the magnetization. The case chosen in Fig. 4, where $M_{fixed}$ and the effective magnetic field are aligned, emphasizes the difference between the torques. In the absence of current, when $M_{free}$ is kicked out of its equilibrium position, it is subject to the "effective-field torque" $T_{field}$, that drives it into precession around the effective field, and the damping torque $T_{damping}$, that brings it back to its equilibrium position. When the current is turned on, $T_{IP}$ is aligned with $T_{damping}$ while $T_{OOP}$ is parallel to $T_{field}$. Depending on the current sign, $T_{IP}$ will then either reinforce the damping or act like an anti-damping. The in-plane torque is therefore useful to stabilize magnetization in its equilibrium position, or, on the contrary, to destabilize it in order to bring it to another equilibrium situation. As for the out-of-plane torque, often called field-like torque, it can emulate the action of a field on $M_{free}$, which means that it can modify the energy landscape seen by the magnetization.

> **There are two contributions to spin torque: the in-plane torque $T_{IP}$ that in some conditions can act like an anti-damping and destabilize magnetization, and the out-of-plane torque $T_{OOP}$, that acts like a magnetic field applied along the fixed layer magnetization.**

### c. Magnetization dynamics with the in-plane spin torque

Because $T_{OOP}$ has long been considered too small to be of use, most spin torque devices are based on the in-plane torque $T_{IP}$ only, as an anti-damping source to destabilize the magnetization without modifying the energy landscape. In that case, as I have contributed to show [22], since magnetization trajectories are constrained by the field-dependent energy profile three different scenarios can arise according to the number of equilibrium positions and their relative stabilities [23] [24]. Fig. 5 illustrates the classical case where the free layer magnetization has two equilibrium positions at zero field, parallel (P state) or antiparallel (AP state) to the fixed magnetization. The device response can be tuned by adjusting the amplitude of the applied field with respect to the coercive field $H_c$, the field required to commute $M_{free}$ between the two stable states.

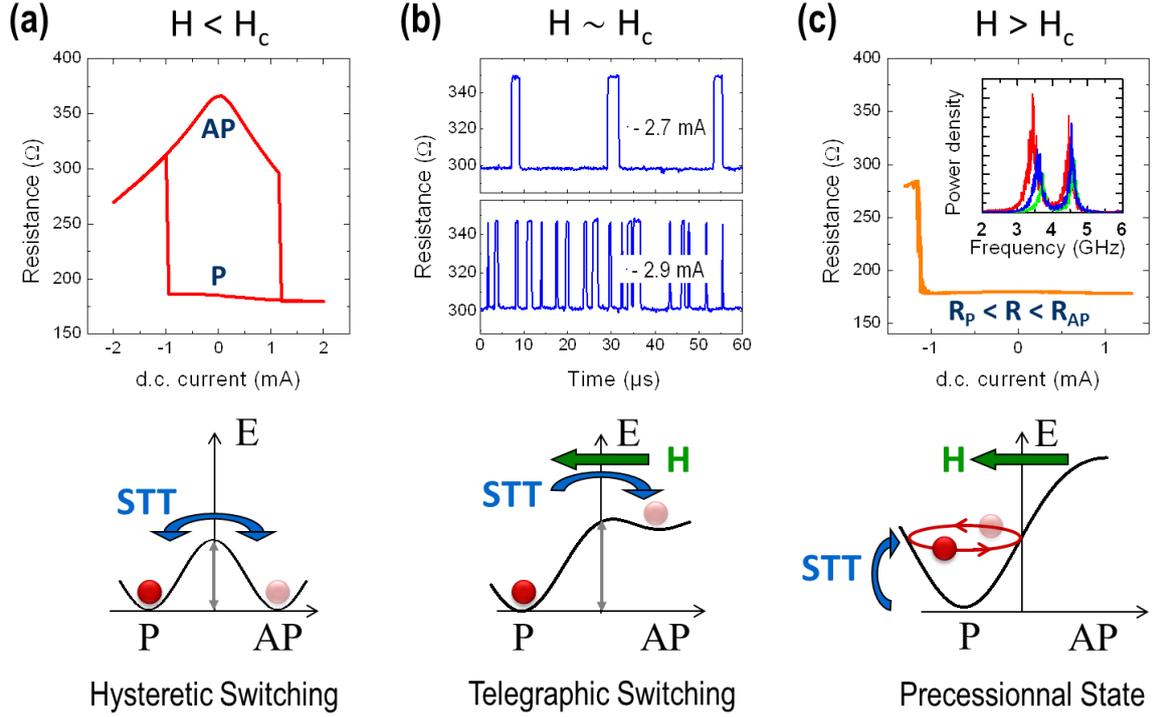

**Figure 5**: Magnetization dynamics scenarios with the in-plane spin torque, as a function of field amplitude. Each case is illustrated with experimental results for magnetic tunnel junctions with an MgO barrier and CoFeB electrodes. (a) $H < H_c$: the spin-torque allows a hysteretic switching between the two stable equilibrium positions. (b) $H \approx H_c$: while spin-torque pushes the magnetization out of its most stable position, the external field destabilizes the second equilibrium position, leading to telegraphic switching. (c) $H > H_c$: only one stable configuration exists but is destabilized by spin-torque, the magnetization goes into precession on a stable orbit.

*1. Hysteretic magnetization switching* (**binary memory**).

At zero or low external magnetic field, lower than $H_c$, both P and AP states are stable (see Fig. 5a). By changing the current sign, the in-plane torque will destabilize either the P or AP state, commuting the magnetization back and forth between these two local energy minima. The free layer magnetization switching is associated with large and sharp resistance variations. The hysteresis loop shows that when the current is turned back to zero the two states remain stable. During my Master thesis we contributed to the first experiments showing this current induced switching at zero field [9]. At that time, experiments were performed in all metallic samples, Co/Cu/Co spin valve nano-pillars made by e-beam lithography in our case. Recently, the optimization of magnesium oxide barriers with low resistivity allows to use magnetic tunnel junctions, which have the advantage of improving the detection of magnetization changes thanks to their high TMR ratio, about 100 % [25] [26], when the GMR in trilayer spin-valves is typically limited to 1-5 %. As shown on Fig. 5a, in that case the bias dependence of TMR leads to the sloped variations of resistance at large currents, more pronounced in the AP state. This spin torque induced magnetization switching at zero field has found a straight forward application in ST-MRAMs and defined a new class of non-volatile binary memories [10].

*2. Telegraphic magnetization switching* (**stochastic device**)

When the applied field approaches $H_c$, stochastic switching between P and AP can occur if the magnetization is destabilized by spin torque in one state, while it is barely stable under thermal fluctuations in the second state [27] [28] [29] [30]. The current amplitude allows modulating the spin torque strength and thus tuning the mean time spent in each state [31], as shown in Fig. 5b. The dwell times might be used to encode probabilities and the current amplitude provides a knob to adjust the odds. This means that Spin Torque can also be used to engineer controlled stochastic devices, for instance random number generators [32].

*3. Sustained magnetization precession* (**microwave nano-oscillator**)
For external fields larger than $H_c$, only one of the states remains stable: the P state in Fig. 5c. When the current is large enough to destabilize the magnetization from the P state, there is no other local energy minima where the magnetization can stabilize. The magnetization then enters a regime of spin torque induced sustained precession [24] [22]. The magnetization orbit is set by the balance between dissipative torques ($\mathbf{T_{damping}}$ and $\mathbf{T_{IP}}$), and conservative torques ($\mathbf{T_{field}}$ and $\mathbf{T_{OOP}}$). In the following, we will focus on this regime of sustained precessional motion.

> **There are three regimes of in-plane spin torque induced dynamics, depending on the amplitude of the applied magnetic field H with respect to the coercive field $H_c$. At low fields ($H < H_c$), the current-induced switching is hysteretic, with applications to binary memories. At intermediate fields ($H \sim H_c$), when the effects of current and field are opposed, the two-level switching acquires a telegraphic character, the current controlling the dwell times of this stochastic device. For large fields ($H > H_c$), only one magnetic state is stable at zero current. If spin torque tends to destabilize it, the magnetization will enter a regime of sustained precession, converted to microwave resistance variations. This is the principle of spin torque nano-oscillators.**

**Related publications:**

J. Grollier, V. Cros, A. Hamzic, J. M. George, H. Jaffrès, A. Fert, G. Faini, J. Ben Youssef, H. LeGall,"Spin-polarized current induced switching in Co/Cu/Co pillars", Appl. Phys. Lett. **78**, 3663 (2001)

J. Grollier, V. Cros, H. Jaffrès, A. Hamzic, J. M. George, G. Faini, J. Ben Youssef, H. LeGall, A. Fert, "Field dependence of magnetization reversal by spin transfer", Phys. Rev. B **67**, 174402 (2003)

## 2. Requirements for applications & first solutions

### a. Requirements for applications

Just as the discovery of GMR boosted data storage in the 1990's, it is foreseen that this latter regime of Spin Torque induced magnetization dynamics can be exploited to build next generation cutting-edge microwave devices for ICT. This new class of microwave nano-devices relies on Spin Torque to induce large amplitude magnetization precessions and magneto-resistance to convert these precessions to electrical signals. The principle of spin torque nano-oscillators is schemed in Fig. 6a. As we have just seen, by dc current injection through the magnetic trilayer, spin torque can induce a sustained precession of the free layer magnetization. This leads to a time varying angle between the free and fixed magnetization layers. Due to GMR or TMR effects, the resistance of the device oscillates, creating an alternating voltage across the junction, as shown for the first time in 2003 [33] [34]. As sketched on Fig. 6b, in the point contact geometry, in addition to the generation of an electrical oscillation, spin waves can be emitted and can propagate since the magnetic free layer is extended [6] [35] [36]. The output voltage frequency is directly in the microwave range, between a few hundred MHz to several tens of GHz [37]. This frequency is linked to the magnetization vibration mode excited by spin torque. It depends on the magnetic material, but also on the geometry. Without modifying the sample shape, the frequency can also be tuned by changing the orbit through a simple variation of the injected dc current amplitude.

These frequency variations are extremely fast, typically a few nanoseconds [38]. Spin torque nano-oscillators are nanoscale, tunable and agile, microwave oscillators. They are extremely promising candidates for diverse applications, in particular telecommunications.

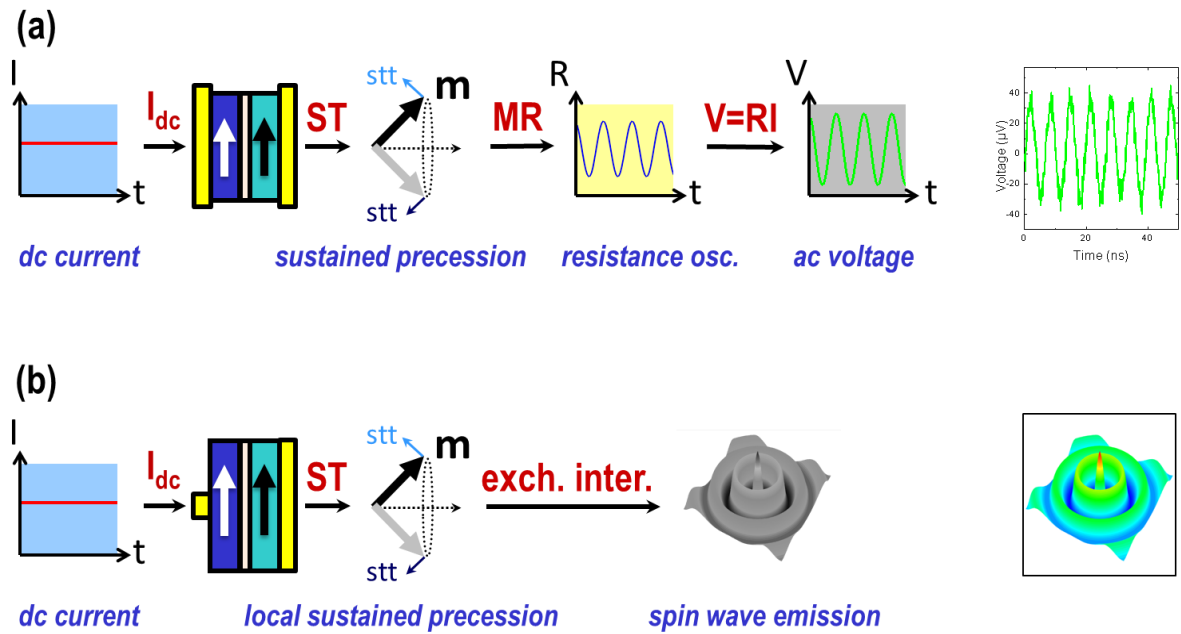

**Figure 6:** Spin transfer nano-oscillators principle (a) microwave oscillator (b) spin wave emitter

These devices are still undergoing intense academic research because a number of issues need to be solved before they can be used in an industrial context. At the first stages of spin transfer nano-oscillators development, in 2003-2005, three main issues appeared. First, having to apply a magnetic field to obtain microwave oscillations was problematic. It cancelled the great advantage of SNTOs: their submicron size. Secondly, the emitted power was far too low, peaking to a hundreds of picowatts [39] [40] [41] for the first GMR based devices. To be able to use STNOs without amplification requires reaching the microwatt range for the less demanding applications. Finally, the spectral linewidth was too large, more than tens of MHz when industrial standards demand less than 1 kHz.

The first point was a motivation for working on the angular dependence of the spin transfer torque, which allowed engineering devices emitting at zero field. The low microwave powers and large linewidths were the drive for studying the non-linear dynamics and synchronization of oscillators.

> **While spin torque microwave devices are extremely promising for telecommunications thanks to their nanometer scale, frequency tunability and radiation-proof character, the first spin torque nano-oscillators were far from meeting the requirements for applications. A large field, problematic for miniaturization was needed to reach the oscillation regime. The microwave power peaked at a few hundred picowatts, when at least microwatts should be emitted. Finally, the spectral linewidth was too large, more than tens of MHz when industrial standards demand less than 1 kHz.**

## b. Zero field emission: engineering the spin transfer torque

The first measurements of microwave emission by spin transfer focused on the uniform vibration mode of magnetization. It was indeed very natural to concentrate on this quasi-macrospin magnetization configuration that gives rise to the well-defined P and AP states required for memory applications. To induce a sustained precession, it is important to create a stable oscillating state. As we have seen, a first possibility is to oppose the effects of applied magnetic field and spin transfer. In the classical "fixed polarizing layer" / normal metal "thin free layer" structures, one current sign favors P state, the positive current in our convention, while the negative current favors the AP state. A strong applied positive field modifies the potential well seen by the magnetization, so that only the P state remains stable. Injecting a negative current destabilizes the only stable state (fixed point) and brings the magnetization towards a precessional trajectory (limit cycle). A first solution to obtain zero field oscillation is to let the fixed layer radiate a dipolar magnetic field on the free layer. In that case, the applied field is zero, but the effective field felt by the free layer is not.

Another solution that we have investigated is to modify the spin accumulation profile giving rise to the spin transfer torque in metallic spin valves, so that one current sign tends to destabilize BOTH P and AP state (while a current of the other sign stabilizes the two states), leading to sustained magnetization precession at zero field. Indeed in metallic spin-valves the spin transfer torque amplitude is proportional to the spin accumulation in the normal metal, at the interface between the normal metal and the free layer [16] [42] [43] [14] (see red line in Fig. 7 and 8). In most studied structures, the fixed layer is chosen as the main spin polarizer, so that when the free layer magnetization reverses the spin accumulation profile is not deeply modified, and in particular its sign remains the same. This situation is illustrated in Fig 7. A direct consequence is that, for these classical trilayer systems, only the current sign determines whether the spin torque favors the P or AP configurations, independently of the free layer magnetization direction.

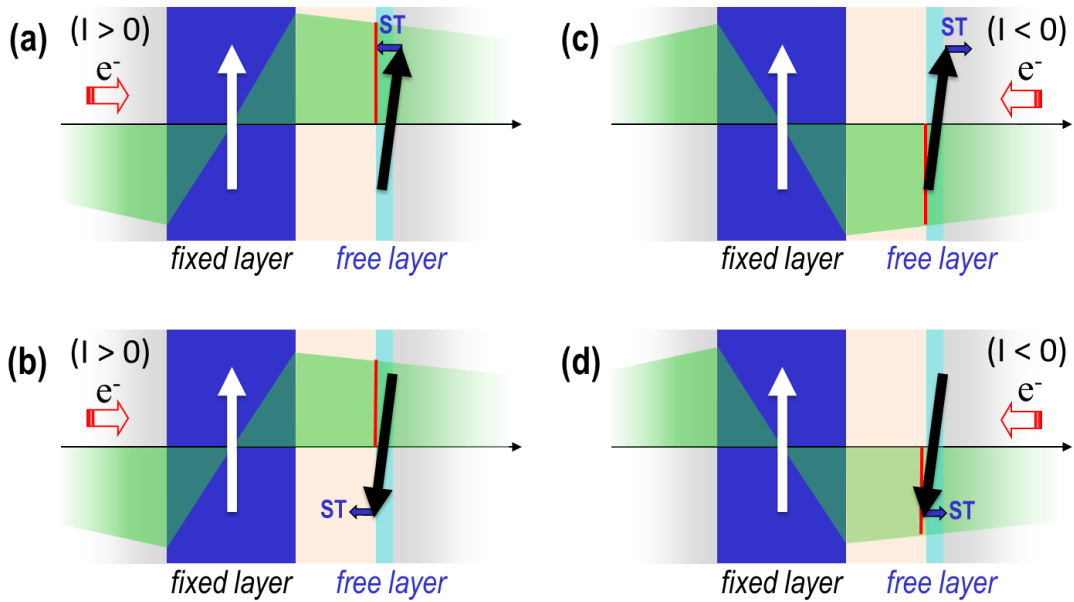

**Figure 7:** Schematic of the spin accumulation profile (in green) and the sign of the torque (given by the spin accumulation –in red– at the normal metal/free layer interface): fixed layer as a polarizer (it is assumed that the free layer does not contribute to the spin accumulation profile) (a) I > 0 favors P state (b) I > 0 destabilizes AP state (c) I < 0 destabilizes P state (d) I < 0 stabilizes AP state

Let's now consider the reverse situation where the free layer is the most efficient spin polarizer, and the fixed layer only slightly influences the spin accumulation profile. This situation is sketched in Fig. 8. This time, when the free layer reverses, the sign of spin accumulation at the interface is modified, and the sign of the torque as well. Therefore, a negative current will stabilize the P and AP configurations, while a positive current will lead to a precessional magnetization state. In these systems, the angular dependence of the torque strongly deviates from the sinus of the angles between magnetizations, and is called "wavy" [44].

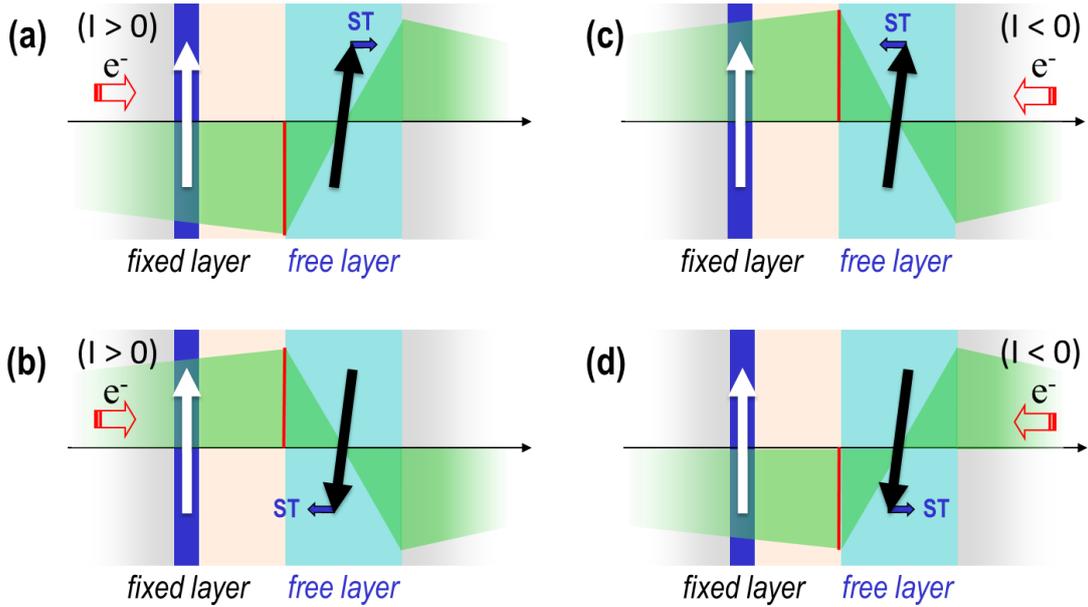

**Figure 8:** Schematic of the spin accumulation profile (in green) and the sign of the torque (given by the spin accumulation –in red– at the normal metal/free layer interface): free layer as a polarizer (it is assumed that the fixed layer does not contribute to the spin accumulation profile) (a) I > 0 destabilizes P state (b) I > 0 destabilizes AP state (c) I < 0 favors P state (d) I < 0 favors AP state

A simple way to achieve this specific configuration is to choose different materials for both layers, with different spin diffusion lengths. The free layer should be a good spin polarizer, so its thickness should be large compared to its spin diffusion length. At the same time it should not be too thick in order to be sensitive to spin torque, which is an interfacial effect. A good material for this purpose is NiFe, which has a small $l_{sf}$ at room temperature, of about 5 nm. The fixed layer should play a minor role in the spin accumulation profile, which can be done by giving it a small thickness compared to its $l_{sf}$. A good choice is Co, which has a spin diffusion length at room temperature of about 60 nm.

In order to test these predictions we have measured such "wavy" Co/Cu/NiFe spin-valves [45] [46]. We have indeed observed spin transfer induced oscillations at zero and very small applied fields, as can be seen on Fig. 9a for a Co(8nm)/Cu(10nm)/NiFe(8nm) trilayer stack. The microwave peak frequency increases with current, in contrast with what is commonly observed for standard trilayer pillars in which the magnetization precesses in the plane of the free layer [33]. We have performed macrospin simulations taking into account the wavy dependence of spin transfer torque. These simulations confirm that an increase of frequency with current is expected for in plane precessions. They also predict an associated decrease of frequency when the in-plane applied field is increased. As can be seen in Fig. 9b, we experimentally observe this trend.

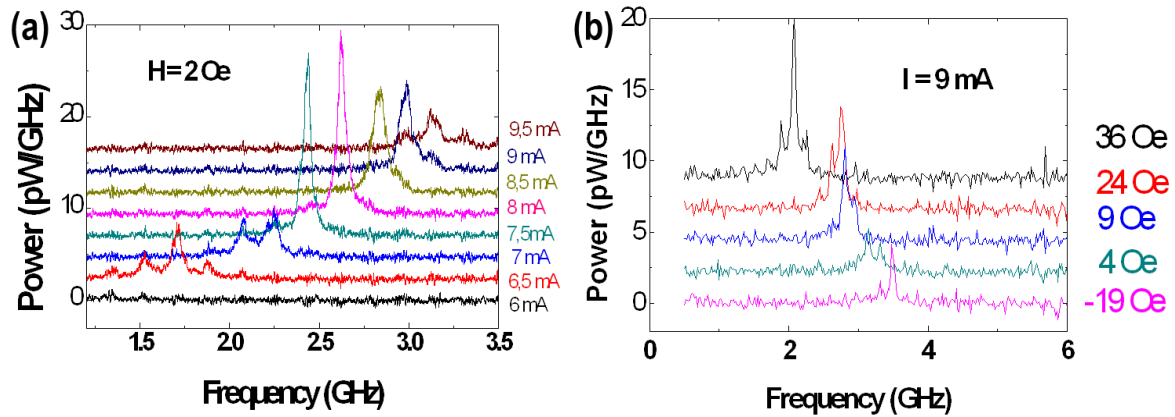

**Figure 9:** Experimental microwave emission for a wavy structure Co 8/Cu 10/ NiFe 8 (in nm). (a) Power spectral density for different injected currents. The applied is zero (H ~ 2 Oe). (b) Power spectral density for different values of the in-plane applied field and a dc current of 9 mA.

In standard trilayers, only a negative current can induce microwave excitations of the free layer magnetization starting from P state. From our previous hand-wavy arguments (see Fig. 8) and more rigorous calculations [44], it appears that it should be the opposite in "wavy" trilayers. Indeed, we have experimentally detected microwave signals only for positive currents. These results have contributed to validate the diffuse calculations of spin transfer torque, which were under debate at that time. In addition they have provided a method for generating zero-field oscillations. As we will see in later chapters, another method to obtain zero field microwave emissions is to replace the quasi-uniform magnetization configuration of the free layer by a vortex configuration.

> **In metallic spin valves, the amplitude of spin torque is set by the spin accumulation at the interface between the normal metal and the free layer. By engineering the spin accumulation profile through a careful choice of fixed and free ferromagnetic materials and their respective spin diffusion lengths, it is possible to create a "wavy" angular dependence of spin torque. In that special case, where the free layer is also the main polarizer in the trilayer, one sign of the current will destabilize both P and AP states, leading to zero-field microwave emission.**

**Related publications:**

**c. A solution to increase power and decrease linewidth: synchronization**

Synchronization is the occurrence of phase locked oscillations of a coupled oscillators assembly with originally dispersed eigen-frequencies. This non-linear effect is very general, and appears in a wide range of systems: biology, chemistry, solid state physics etc. Indeed, the same phenomena rules the synchronization of neurons spiking, crickets chirping, fireflies glowing, pendulum clocks beating and Josephson junctions oscillating. As pointed out by Steven Strogatz in his book "Sync", synchronization is an astonishing phenomenon in a world where we are more accustomed to increasing entropy and chaos than a sudden emergence of order [47]. In biology, synchronization can be interpreted as a trick developed by nature to enhance communication. Indeed, when an oscillator assembly is synchronized, the total emitted power is much larger than the power emitted by a single isolated oscillator, and signal fluctuations are reduced. For the gathering of male crickets or fireflies, synchronization is a strategy to attract more efficiently the females flying by, through an increased sound or light. In the brain, where neural spike timing could be at the heart of the information code, neural synchronization seems to be one of the keys of information processing, by adjusting the phase of neighboring neurons, or even neurons of different brain areas. In particular, synchronization of local neural assemblies, by raising the signal amplitude, facilitates the communication between the different brain regions [48].

If the total number of oscillators is N, the total power can be increased as much as $N^2$ and the emission linewidth as $1/N$ [49]. Synchronizing an assembly of spin transfer nano-oscillators can therefore be a good strategy to concomitantly enhance the emitted power and reduce the linewidth. To obtain the phase locked oscillations, the oscillators have to share the information about their respective phases, and be able to influence the other's phases, i.e. they have to be coupled. In 2005 it has been demonstrated that two spin transfer oscillators can be coupled through the interaction of their emitted spin waves [50] [51]. In that case the two oscillators were neighboring nanocontacts, about 500 nm apart, sharing their free layer. While this coupling scheme appears quite efficient, it can only be used to decrease the emission linewidth, but not to increase the power. Indeed, the particular coupling geometry requires the nanocontacts to be connected in parallel. This shunts the total emitted power so that, even if the oscillators are perfectly synchronized, the final power is equal or less than the emission of a single nanocontact. In addition, this mechanism is local, as the interaction is strong only between neighbouring contacts (spin waves in 3d ferromagnetic metals typically propagate over a few microns [35] [36]), and might not be efficient to synchronize an assembly of several oscillators.

On the other hand, a system very close to the spin transfer oscillator: Josephson junctions, can be efficiently synchronized by a global microwave coupling when electrically connected [52]. Plus, as demonstrated by the NIST team in Boulder, a spin transfer oscillator can be locked to the frequency of an injected microwave current [53]. These considerations led me to propose another coupling scheme, this one global, simply based on the electrical connection of the oscillators. In a first approach, I performed macrospin numerical simulations to compute the dynamics of spin transfer oscillators assemblies, electrically connected in series (as illustrated on Fig. 10a), or in parallel, and to identify the conditions for synchronization.

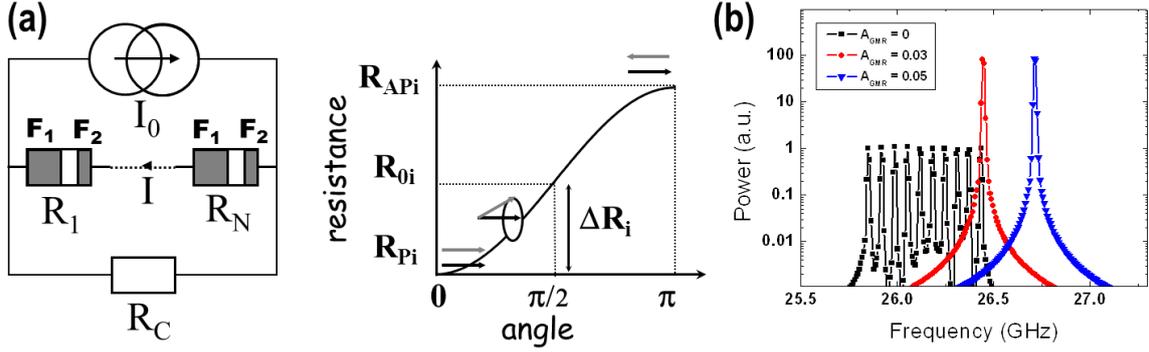

**Figure 10:** Macrospin simulations of electrically coupled STNOs. (a) Left: Schematic of N oscillators connected in series and coupled to a load resistance $R_c$. Right: resistance variation as a function of the angle $\theta_i$ between the fixed and free magnetizations of each oscillator. (b) Logarithm of the normalized emitted power for 10 oscillators and different values of magnetoresistance.

In arrays of electrically connected oscillators, each oscillator is biased by a dc current plus a microwave contribution created by the resistance oscillations of all the other oscillators [54]. This means that the larger the magneto-resistance (MR effect), the larger the coupling. In these calculations, we have found that very modest values of MR (3 % for the oscillators considered in Fig. 10b) can allow a full synchronization for experimentally achievable frequency dispersions. It is important to mention that these calculations are done at zero temperature, which means that the linewidth of each oscillator is zero. In addition, the oscillator tunabilities obtained within the macrospin model are much larger than the ones observed in real samples. We will see in the following that these two factors: linewidth and tunability have in fact a strong impact on the ability of spin transfer oscillators to lock, which complicates their synchronization.

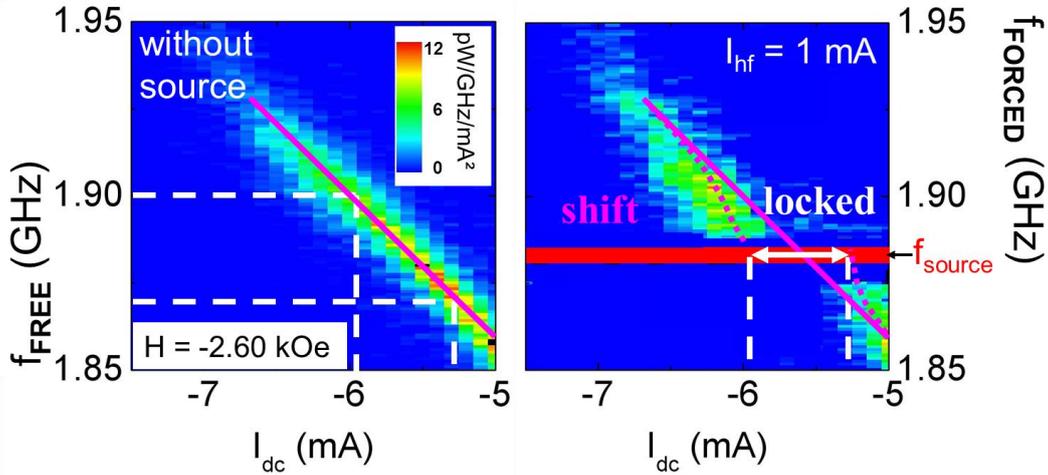

**Figure 11:** Frequencies of the free (left) and forced (right) oscillator as a function of the injected current. The color scale gives the power spectral density. The frequency of the external source is 1.88 GHz, and the microwave current amplitude is set to 1 mA.

In order to experimentally determine the coupling amplitude by this electrical mean, we have first performed phase-locking experiments of a spin-valve spin transfer oscillator to a microwave source. Fig. 11 shows the frequency evolutions of the free and forced oscillator as a function of the injected dc current. As can be seen on the graph on the right, there is a frequency range in which the frequency of the forced oscillator is equal to the frequency of the source. Out of this locking range, the oscillator frequency is pulled towards the source frequency: $f_{Forced}$ is different from $f_{Free}$. We have then focused on the impact of the intrinsic microwave properties of the spin transfer oscillator, linewidth and tunability, on its ability to phase lock to a microwave source.

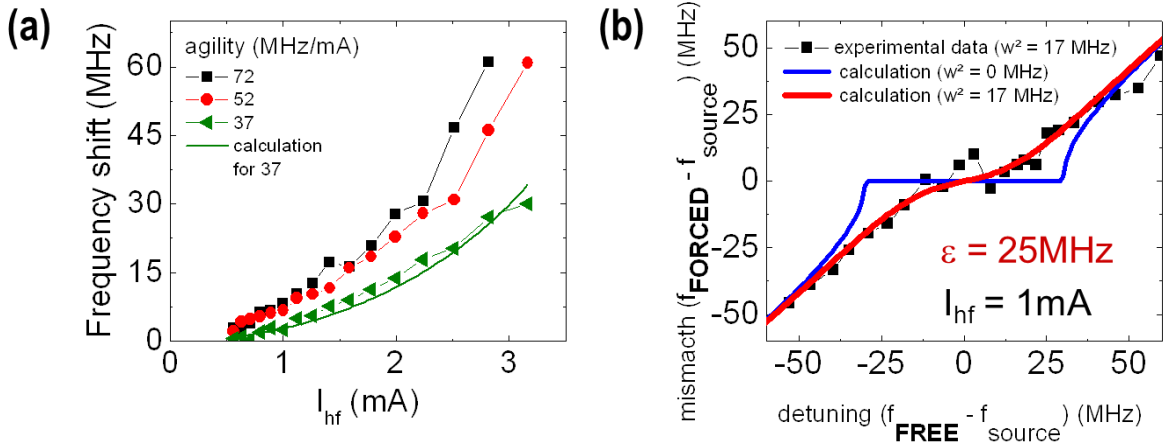

**Figure 12:** (a) Evolution of the frequency pulling as a function of the injected dc current, for different values of the tunability. The line is a calculation for a tunability of 37 MHz/mA. (b) Frequency mismatch versus frequency detuning. The black dots are the experimental data. The lines are a calculation for zero linewidth (blue) and 17 MHz corresponding to the experimental value (red).

Fig. 12a shows the impact of tunability on the ability of the oscillator to lock: the frequency shift due to the microwave source pulling increases with increasing tunability. This is not unexpected. Indeed, the tunability of an oscillator defines its ability to change its frequency in response to a current variation. In other words and this is a well-known fact in synchronization theory: the larger the oscillators' non-linearity, the better they can synchronize. We have also experimentally demonstrated that the spectral linewidth of the oscillator has a strong impact on the locking range. Indeed, a large part of the linewidth originates from phase fluctuations [55]. Since synchronization is the process of finely adjusting the phase to an external signal, a very fluctuating phase (i.e. a large linewidth) is detrimental for synchronization. Fig. 12b shows an experimental result (black squares) giving the frequency mismatch ($f_{Forced} - f_{Source}$) versus the detuning ($f_{Free} - f_{Source}$) between the oscillator and the source for an oscillator with a finite linewidth of 17 MHz. As it appears immediately from the data, the locking range, i.e. the frequency range where the mismatch is zero, is extremely small.

To interpret this data, starting from the equation for magnetization dynamics including the spin torque due to the microwave source [56], and following Adler's formalism [57] [49], we have first calculated the relevant uniformly rotating phase Φ for the oscillator, that depends on the non-linearity, to end up with the equation driving the phase dynamics of the spin transfer oscillator in the presence of noise:

$$\frac{d(\Delta\Phi)}{dt} = 2\pi(f_{Free} - f_{Source}) + \varepsilon \sin(\Delta\Phi) + \xi(t)$$

In this equation, ΔΦ is the phase difference between the oscillator and the source, ξ(t) is the noise amplitude, and ε is the coupling strength. The oscillator is synchronized if ΔΦ is constant. In a first step, this equation has allowed us to fit the experimental data with the coupling strength ε, using the experimental linewidth values to account for noise. As can be seen from the fit results given by lines in Fig. 12 the agreement is very good. In Fig. 12b we show the results of two calculations: the first one in blue done by setting the linewidth to zero, the second one, in red using the real experimental value. The comparison between the two results strikingly demonstrates the importance of using small linewidth devices to obtain large locking ranges. Another important output of this calculation is the expression of the coupling strength as a function of the microwave characteristics of the oscillator:

$$\varepsilon = \frac{I_{hf}}{2\sqrt{2}} \sigma \tan(\gamma) \sqrt{\frac{I_{dc}}{I_{dc} - I_{th}}} \sqrt{1 + \left(\frac{2\pi}{\sigma} \frac{I_{dc}}{I_{th}} \frac{\partial f_{Free}}{\partial I_{dc}}\right)^2}$$

This formula is in agreement with earlier calculations of the locking range of spin transfer oscillators performed by A. Slavin et al. using a different approach [58]. In this equation, $I_{hf}$ is the amplitude of the injected microwave current, $\sigma$ the spin torque efficiency, $I_{dc}$ the dc current, $I_{th}$ the threshold current for auto-oscillations, $\gamma$ the equilibrium angle between the fixed and free magnetizations and $\partial f_{Free}/\partial I_{dc}$ the tunability.

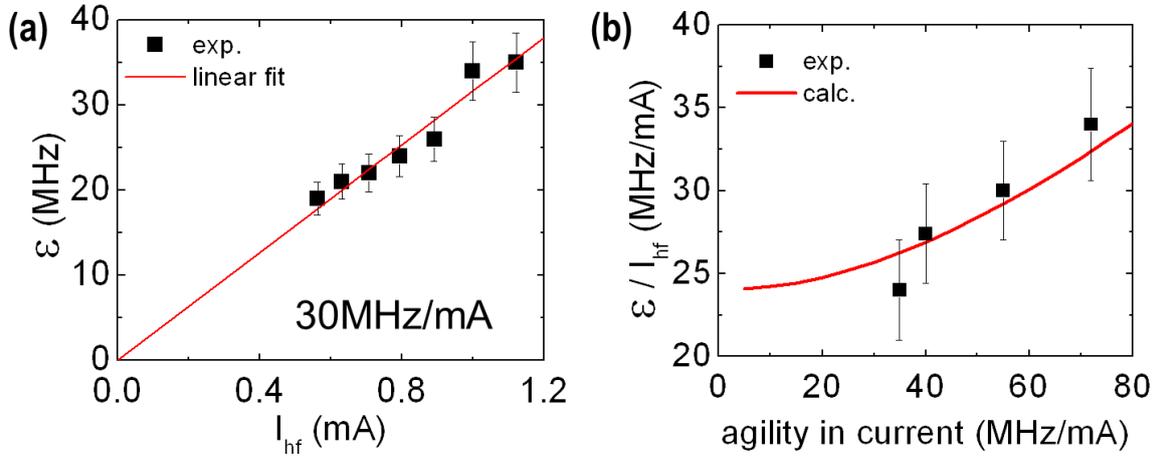

**Figure 13:** (a) Variation of the coupling strength $\varepsilon$ versus the microwave current $I_{hf}$ for H = 2.65 kOe, $I_{th}$ = - 3 mA, $I_{dc}$ = - 5 to - 8 mA, $\partial f_0/\partial I_{dc}$ = 72 MHz/mA, $w^2$ = 17 MHz. The experimental points (black dots) are the best fits to the curve mismatch vs detuning for different values of $I_{hf}$, with noise corresponding to the experimental linewidth of 17 MHz (b) Black squares: experimental variation of $\varepsilon/I_{hf}$ as a function of the agility in current for our STNO such as $I_{th}$ = - 3 mA, $I_{dc}$ = -6 mA, $w^2$ = 13 MHz. Red line: calculations taking $\sigma$ = 1 GHz/mA and $\gamma$ = 2.75°.

To test this expression, we have studied how the coupling strength, extracted from the experiments thanks to fits similar to the ones of Fig. 12, evolves with $I_{hf}$ and the tunability. These experimental results are shown on Fig. 13, together with the theoretical curves obtained with the calculated value of the coupling strength $\varepsilon$. The quantitative agreement demonstrates the validity of the analytical approach [59].

We have then extended our analytical calculations to the case of electrically connected oscillators arrays, in series and/or in parallel [60]. We have found that the equation for the phase dynamics of each coupled oscillator was similar to the very general Kuramoto's equation [61], which has an analytical solution. Using the previous coupling strength formula validated by our experimental results, we have predicted the condition to synchronize an array of electrically connected spin transfer oscillators:

$$\frac{\Delta R_{osc}}{R} > \frac{2\,(LW+D)}{I_{dc}} \frac{1}{\left(\varepsilon/I_{hf}\right)}$$

$\Delta R_{osc}/R$ is the fraction of magnetoresistance converted to oscillations, LW the spectral linewidth and D the frequency dispersion (the oscillators are considered identical except for their dispersed eigenfrequencies). A practical consequence of this result is that to synchronize the assembly, each oscillator already needs to have good microwave characteristics: a small linewidth, a large tunability, and a large emitted power (in other words, a large magnetoresistance). If we set the value of $\varepsilon/I_{hf}$ to the experimental value of 30 MHz/mA (see Fig. 13), the dc current to 5 mA, the linewidth to 20 MHz and the dispersion to 100 MHz, we find that for synchronization $\Delta R_{osc}/R$ should be larger than 5 %. If we consider that 20 % of the total magnetoresistance is effectively converted to oscillators, this means that the oscillator array can synchronize if the MR of each oscillator is larger than 25 %. If the requirements of large tunability, small linewidth and small frequency are common to all coupling schemes, it appears that synchronization through the emitted microwave currents requires devices with a magnetoresistance much higher than can be obtained in metallic spin-valves. Magnetic tunnel junctions are therefore very interesting candidates for synchronization through electrical coupling.

**Synchronizing an assembly of oscillators is a possible solution to increase the emitted power and the spectral purity. To obtain these phase locked oscillations, a coupling between oscillators is needed. While coupling of two nano-contact oscillators through the interaction of their emitted spin waves has been experimentally demonstrated, this method has two drawbacks. First, since the oscillators share their free layer, they have to be connected in parallel, which shunts the emitted power and prevents any gain on this side (this method can nevertheless be used to increase the spectral purity). Secondly, spin wave interaction is a neighbor to neighbor coupling, which can be detrimental for the synchronization of several oscillators. This is why we focused on the synchronization of serially connected oscillators through their emitted microwave currents. Thanks to macrospin simulations we have shown the possibility to synchronize several oscillators by this method. We have then performed phase locking experiments of single oscillators with quasi-uniform magnetization to a microwave source. The results, interpreted thanks to analytical modeling of the phase dynamics, show the importance of two factors for synchronization: a large tunability, as a hint of the oscillator faculty to adapt its frequency, will facilitate the coupling, while a large linewidth, indicating a strong phase noise, will be detrimental to phase locking. By extending the model to an array of electrically connected oscillators, we have derived the conditions for synchronization. In addition to strong tunability, low phase noise, and low frequency dispersion between oscillators, synchronization will require each oscillator to emit a large power, larger than possible with spin-valve nano-oscillators due to their magneto-resistance ratios below a few %.**

## 3. Increasing the microwave power: magnetic tunnel junctions

### a. Quasi-uniform precessions in magnetic tunnel junctions

The recent developments of magnetic tunnel junctions with a low resistivity MgO tunnel barrier have allowed the injection of the high current densities required to manipulate the magnetic configuration through spin torque in these structures [25] [26]. Studying the sustained magnetization precessions in these magnetic tunnel junctions is very interesting since the emitted microwave power evolves as the square of magnetoresistance, typically 100 % at room temperature in these devices (to be compared to a few % for metallic spin-valves). Spin torque induced microwave emissions up to the microwatt have been measured with this kind of samples [62] [39]. These MgO based magnetic tunnel junctions are therefore extremely interesting candidates for synchronization. Unfortunately, the first measurements, focused on studies of the quasi-uniform precessional mode, have also highlighted a significant drawback of MgO junctions microwave characteristics. As it appears clearly on Fig. 14a showing a typical power spectra, the large microwave emissions are associated with very large linewidths, above 100 MHz, and up to a few GHz. Our study of the microwave emission of Fe/MgO/Fe junctions reveals that the linewidth in these epitaxial stacks is as large as in CoFeB/MgO/CoFeB junctions, indicating that the textured nature of these latter is not at the origin of the poor quality factors [63]. As we have seen in the previous section, large linewidths, correlated to large phase noise, are extremely detrimental for synchronization.

To gain insight into the origin of this degraded spectral purity, we have studied the emission linewidth of PtMn 15/ CoFe 2.5 / Ru 0.85 / CoFeB 3 / MgO 1.075 / CoFeB 2 (nm) junctions patterned down to the anti-ferromagnet in the shape of a 170x70 nm$^2$ ellipse. The TMR ratio, 100% at 300 K, increases to 140% at 20 K.

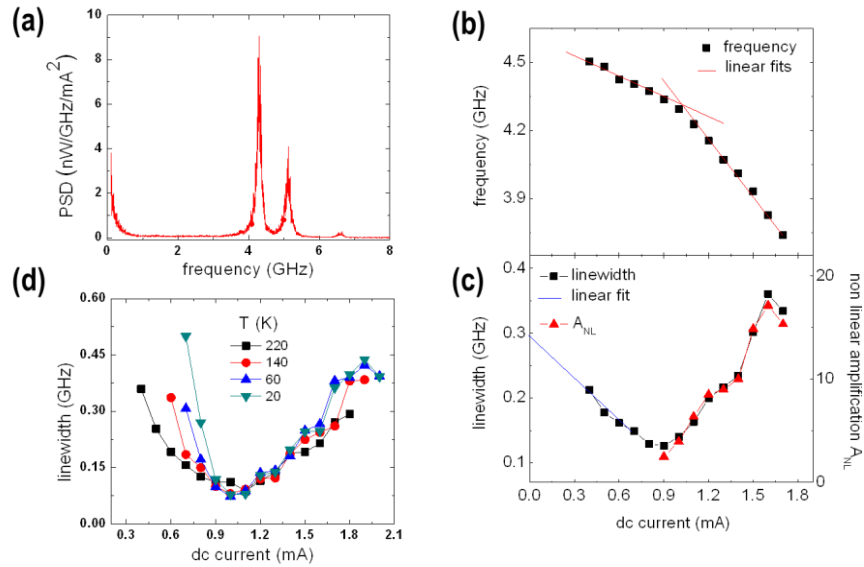

**Figure 14:** (a) Representative power spectral density corresponding to the quasi-uniform vibration mode of a CofeB/MgO/CoFeB magnetic tunnel junction, obtained for pour $I_{dc}$=1 mA and an in-plane easy axis field H =110 Oe at 300 K. Two large peaks are observed. (b) Variation of the LF peak frequency as a function of dc current (black squares) for H = 110 Oe at 300 K. The lines are linear fits. (c) Left axis: the black squares give the evolution of the LF mode linewidth as a function of $I_{dc}$. Right axis : variation of the non-linear amplification factor ANL with $I_{dc}$ (red triangles). (H = 110 Oe at 300 K) (d) Linewidth as a function of dc current, for T =20, 180 and 300 K, and H=205 Oe.

In the theoretical framework established by J.-V. Kim, A. Slavin and V. Tiberkevich [64] [65] [55], there are two regimes with distinct properties as far as the linewidth is concerned. In the sub-critical regime, i.e. below the critical current for the generation of self-sustained oscillations, the spin transfer gradually compensates the damping when the current is increased. The linewidth Δf can then be written as: $\Delta f = \Gamma_g - \frac{\sigma}{2\pi} I_{dc}$, where σ characterizes the spin transfer efficiency, and $\Gamma_g = \alpha \frac{\gamma \mu_0 M_{eff}}{2\pi}$ is the classical FMR linewidth for in-plane applied fields. The experimental variation of the linewidth with current is shown on Fig. 16c. In the sub-critical range from 0 to about 0.9 mA, the linewidth decreases linearly with current according to predictions, and a fit with the previous expression gives a reasonable value of 0.009 for CoFeB (see blue line in Fig. 16c).

In the over-critical regime of sustained magnetization oscillations, the spin torque perfectly compensates damping, and there are two main contributions to the linewidth:
$$\Delta f = A_{NL} \times \Gamma_g \frac{P_n}{E(p_0)}.$$
The first contribution, $A_{NL}$, arises from the non-linearity of spin torque nano-oscillators. The more is oscillator is non-linear, the more its frequency becomes sensitive to external fluctuations, the larger the phase noise. The non-linearity of the oscillator can be quantified by its tunability $df/dI_{dc}$. The non-linearity parameter can indeed be rewritten as:
$$A_{NL} = 1 + \left(\frac{I_{dc}}{\Gamma_g} \frac{df}{dI_{dc}}\right)^2.$$
The experimental frequency variation with current is given in Fig. 14b. From this curve, it is possible to calculate $A_{NL}$ (red triangles in Fig. 14c) and compare it to the linewidth evolution with current (black squares in Fig. 14c). In the overcritical regime, the agreement is excellent, which validates the large contribution of the non-linearity to the linewidth pointed out and introduced via $A_{NL}$ in the model. We can note here that synchronization always requires to compromise between large tunability (non-linearity) and low linewidth, which are hardly compatible.

The second contribution to the linewidth in the model arises from thermal fluctuations (which are then amplified by the non-linearity): $P_n = k_B T$ is the thermal noise amplitude, and $E(p_0)$ the oscillator energy. In this framework, a large decrease of linewidth with temperature is expected. Fig. 14d shows the evolution of linewidth with temperature. It appears clearly that in the over-critical regime, for currents above 0.9 mA approximately, the linewidth is practically independent on temperature. Additional measurements allowed us to understand that, while the model remains valid, the noise to take into account in $P_n$ is not just thermal noise, but a background noise due to a spin torque induced chaotization of the magnetic system [66].

The origins of this chaotic dynamics of the quasi-uniform magnetization mode in magnetic tunnel junctions are still poorly understood. Similar metallic systems, in terms of materials, dimensions and stack (including a Synthetic Anti-Ferromagnet as well) show much smaller linewidths, down to 4 MHz [67]. Nevertheless, even in metallic spin-valves the experimental linewidths stay finite at low temperature, indicating a remaining source of noise at 0 K [68]. As can be seen in Fig. 14a, the typical spectra for MgO tunnel junctions with an in-plane field shows two peaks. These two modes have been respectively attributed to a center and an edge mode resulting from the dipolar coupling with the reference layers in the SAF [39] [69]. The large spectral power at low frequency, increasing when the frequency approaches zero, indicates that the magnetization, due to thermal activation, is constantly changing orbit between the two modes, generating this telegraph like spectra. It could be expected that when the temperature decreases, the mode hopping disappears, giving rise to a single peak spectra.

This is absolutely not the case. As shown in Fig. 15, the two peaks remain present. In addition the temperature dependence of the power spectral density is complex, and not monotonic. This cannot be explained by the TMR variations, experimentally linear with temperature.

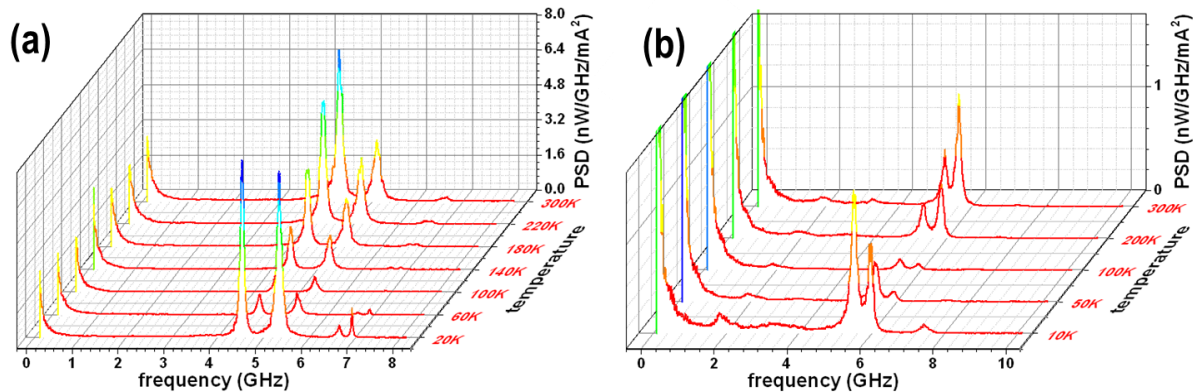

**Figure 15:** Power spectral densities as a function of temperature for two different junctions, measured under different bias conditions (a) Junction 1, Idc = 1 mA, H = 110 Oe (P state) (b) Junction 2, H = -260G, Idc = -1.6 mA (AP state)

One of the major differences between metallic systems and magnetic tunnel junctions are the amplitudes of injected dc currents that differ from one order of magnitude, leading to very different amplitudes of the current-induced Oersted field. It might then be that the Oersted field has a major impact on spectral purity. Large at the edges, it will indeed tend to kill the edge mode, and stabilize the dynamic center mode. This interpretation is coherent with the general observation of single peaks in spin valves, even when the stack involves a SAF [67].

It would be therefore interesting to engineer the magnetic tunnel junctions in order to favor the appearance of a single mode, well separated in energy from other modes, to avoid spin torque induced mode hopping and chaotization, and finally obtain low linewidths combined with large microwave powers. This is the topic of next section.

> **MgO barrier based magnetic tunnel junctions, with their high MR ratios (~ 100 %) and low resistivity, are good candidates for synchronization. The first measurements of microwave emissions in MgO tunnel junctions have indeed demonstrated emitted power in the microwatt range, close to the requirements for the less demanding applications. Nevertheless a great disappointment came from the related spectral purity, with very large linewidth over 100 MHz, unfit for applications, where less than 10 kHz is the norm, and unfit for synchronization. For that reason we have experimentally studied the evolution of the emission linewidth of MgO based MTJs as a function of current and temperature. We have confirmed the predicted large contribution of non-linearity to linewidth. We have also shown that temperature alone cannot account for the linewidth. In particular the latter stays finite at low temperatures. Our results indicate a spin torque induced chaotization of the magnetic system, leading to strong mode-hopping even at low temperature.**

**Related publications:**

### b. Vortex Gyration in magnetic tunnel junctions

As just mentioned, the first measurements of microwave emission in magnetic tunnel junctions have been promising from the point of view of the emitted power, but very disappointing in terms of spectral purity. In parallel to these investigations, several teams started to study vortex induced spin torque oscillations in metallic spin-valves [70] [71] [72]. As can be seen on Fig. 16b the associated microwave emissions have very thin linewidths, a few hundreds of kHz, much smaller than those corresponding to the quasi-uniform mode in magnetic tunnel junctions, recalled in Fig.16a. This feature can be easily understood: the lower energy mode of vortex vibration, the gyrotropic mode of the core, corresponding to a circular trajectory of the core around its equilibrium position, is very far in energy (a few GHz) from the other vibrations modes, related to the vortex "wings" and not its core. Of course, since these results were obtained in spin-valves, the total output power in Fig16b, which can be roughly estimated by multiplying linewidth and maximum power density, is much lower than for the tunnel junction in Fig. 16a.

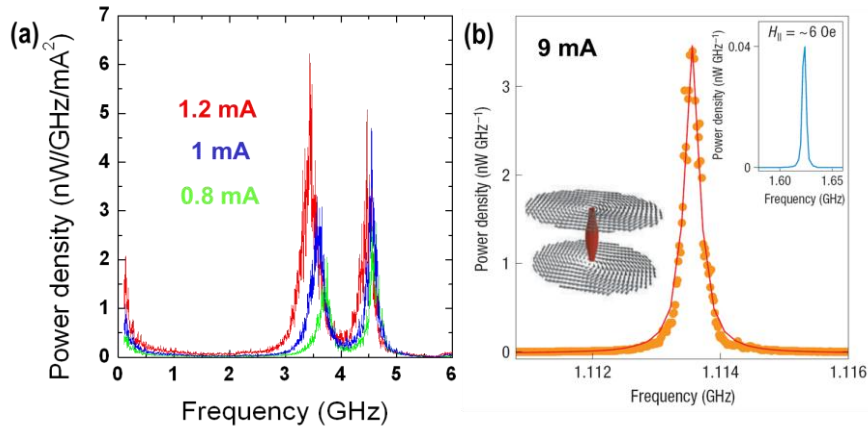

**Figure 16:** (a) Microwave spectra obtained by spin torque excitation of the quasi-uniform mode in a CoFeB/MgO/CoFeB magnetic tunnel junction ($H_{perp}$ = 595 Oe) (b) Adapted from ref. [70] Microwave emission corresponding to the excitation of a vortex mode in a Co/Cu/Co spin valve (H = 480 Oe).

In order to keep a good spectral purity while maintaining a large power, we then decided to introduce a vortex in the free layer of an MgO based magnetic tunnel junction. This can be achieved simply by using a large circular junction to allow the in-plane magnetization to curl, and a thick enough free layer to allow the vortex core to pop out. Fig. 17a and b give a schematic of our structure. The free layer containing the vortex is a thick 15 nm NiFe layer with a diameter of 170 nm, while the polarizing layer is the top CoFeB layer of the synthetic antiferromagnet base of the junction. The results are shown on Fig. 17c. By applying an out-of-plane field (5.1 kOe in the case of Fig. 17c), microwave oscillations are obtained, with a linewidth of about 1 MHz, much lower than observed for the uniform mode of Fig. 16a.

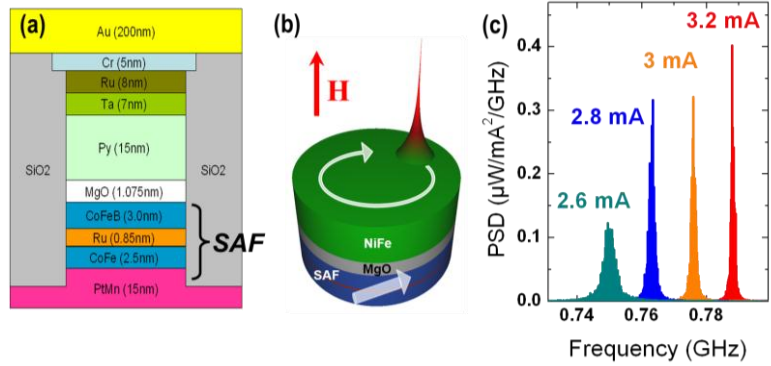

**Figure 17:** (a) Schematic of the SAF/MgO/NiFe magnetic tunnel junction stack. (b) Schematic of the structure with a vortex in the free layer. A field applied perpendicular to the plane allows lifting the polarizer out-of-plane. (c) Microwave spectra measured for a 5.1 kOe out-of-plane magnetic field.

The resulting frequency, lower than the GHz, is typical from the gyrotropic mode of the vortex core. The integrated power, close to 5 nW, is largely improved compared to spin-valves. It is smaller than for CoFeB/MgO/CoFeB junctions: the TMR is indeed limited in these samples to 15 %, due to the MgO/NiFe interface that kills the filtering effects at the origins of large TMR ratios. We will see in the following that inserting 1 nm of CoFeB between MgO and NiFe allows recovering the large TMR values. The evolution of integrated power as a function of the applied field is shown on Fig. 18a. At zero field, the oscillator does not emit, but an out-of-plane field of a few kOe allows spin torque induced vortex core gyrotropic oscillations. At larger fields, the power decreases again when the vortex transforms to a uniform out-of-plane magnetization configuration.

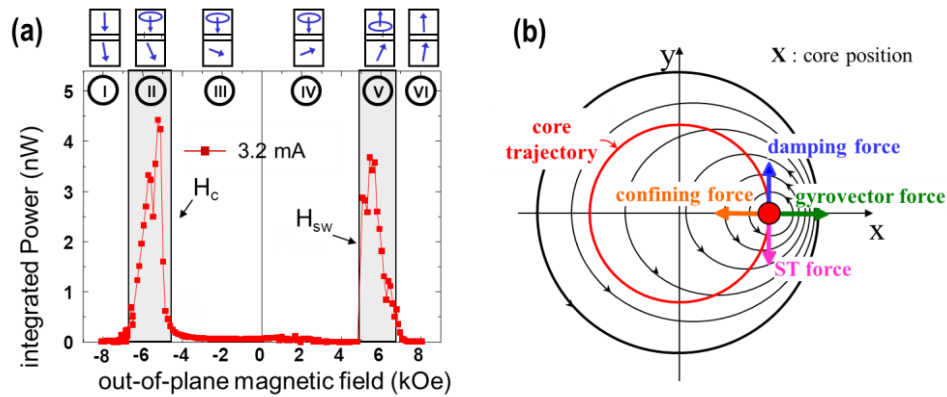

**Figure 18:** (a) Evolution of integrated power as a function of the out-of-plane applied field for the SAF/MgO/NiFe oscillator. (b) Schematic of the forces acting on the vortex core: confinement, gyrovector, damping and spin torque.

The SAF/MgO/Vortex tunnel junctions are a model system to study spin torque induced vortex dynamics as the synthetic antiferromagnet maintains the polarizing layer fixed and uniform. It is necessary to take into account the non-uniform distribution of magnetization in the vortex to obtain the effective spin torque force acting on this magnetic object. I have analytically calculated the spin transfer force exerted on a vortex by a fixed and uniform polarizer. The calculation is based on the assumption that the vortex shape is rigid (two vortex ansatz [73]), and that the vortex core has a perfectly circular orbit around the center of the dot. The energy given by spin torque to the vortex is integrated over the volume of the dot for a given orbit radius. The same procedure can be applied to the energy dissipated due to the damping; equaling these two quantities corresponds to the assumption of stationary sustained oscillations. The resulting expression for the spin transfer force is: $\vec{F}_{ST}^{\perp} = j p_z \sigma \pi M_s (\vec{e}_z \times \vec{X})$, where j is the current density, $p_z$ the out-of-plane component of spin polarization, σ the spin transfer efficiency, $M_s$ the saturation magnetization of the free layer, **e_z** the out-of-plane unit vector, and **X** the in-plane vortex core shift. We have validated this expression by showing that the critical current densities it predicts for sustained vortex precessions are extremely close to the values obtained by micromagnetic simulations [74]. It should be noted that previous calculations had allowed deriving a quite similar expression, unfortunately erroneous of a factor 2, due to a bit too rapid extension of Thiele approach to the calculation of the spin transfer force [75]. Later works, based on a correct development of Thiele equation, concluded to the same expression as ours [76].

The expression we have obtained for the spin transfer force acting on a vortex when the polarizer is fixed and uniform allows a first conclusion: if the out-of-plane component of spin polarization is zero, then $F_{ST}$ is also zero. This explains the observations of Fig. 18a: a strong out-of-plane field is needed for $p_z$ amplitude to be large enough for the spin torque to win over damping. It is possible to go further and determine the exact conditions to obtain spin transfer induced vortex precessions. Fig. 18b illustrates the vortex core circular trajectory (in red), as well as the different forces at stake. Radially, two conservative forces are opposed: the confinement force (magnetostatic charges plus Oersted field created by the current injected perpendicularly to the plane) and the centrifugal gyrovector force arising from the particular vortex geometry. Tangential to the trajectory, two dissipative forces are fighting: the spin transfer force, and the damping force, the latter always opposed to the speed. The speed vector direction is fixed by the vortex core polarity P. Consequently, according to the value of P, +1 or -1, corresponding to the core magnetization pointing up or down, the damping force will point in different directions. The core polarity will therefore play a role in the condition to obtain oscillations, the spin transfer force having to be opposed to the damping force. From these considerations and the expression of **F**$_{ST}$, it can finally be deduced that the conditions to observe microwave emissions via vortex precession are: $j > j_c$ and $j.P.p_z > 0$ (here $j > 0$ corresponds to electrons flowing from the free to the polarizing layer). We have validated these conditions both by micromagnetic simulations and experiments [41].

**To prevent the loss of spectral purity through mode-hopping, we have turned to the study of vortex oscillations in MgO-MTJs. Indeed the low frequency gyration of the magnetic vortex core is far away in energy from the other eigen-modes, related to excitations of the vortex "wings". This has allowed us to obtain experimentally low linewidths, of the order of 1 MHz, together with large emissions of a few nanowatts , for perpendicularly applied magnetic fields of a few kOe. In our magnetic tunnel junctions, the polarizing layer is the top layer of a strong synthetic antiferromagnet stack, so it is fixed and uniform. For that reason it is a model system for deriving the spin transfer force acting on the non-uniform magnetic vortex. We have done so by calculating the energy gained via spin torque during one vortex rotation over the volume of the dot. The resulting spin transfer force induced by a fixed uniform polarizer on a vortex is proportional to the out-of-plane component of the polarization. In other words, if the polarizing layer magnetization naturally lies in the plane as it is the case in our junctions, the application of a perpendicular field is necessary to lift it out-of-plane. That is why, in our [SAF fixed uniform polarizer] / MgO/ [NiFe free layer], a perpendicular field is necessary to obtain spin torque induced vortex oscillations.**

**Related publications:**

A.V. Khvalkovskiy, J. Grollier, A. Dussaux, K. A. Zvezdin, V. Cros, "Vortex oscillations induced by spin-polarized current in a magnetic nanopillar : analytical versus micromagnetic calculations", Phys. Rev. B **80**, 140401 (R) (2009)

A. Dussaux, B. Georges, J. Grollier, V. Cros, A.V. Khvalkovskiy, A. Fukushima, M. Konoto, H. Kubota, K. Yakushiji, S. Yuasa, K.A. Zvezdin, K. Ando, A. Fert, "Large microwave generation from current-driven magnetic vortex oscillators in magnetic tunnel junctions", Nature Com. **1**, 1 (2010)

## 4. Vortex Oscillations at zero field

### a. Perpendicular polarizer

We have succeeded to engineer highly coherent vortex oscillations with a large power, but unfortunately, under large applied fields, the spin transfer force being proportional to $p_z$ as we have just seen. A solution to obtain zero field oscillations is then of course to use a perpendicular polarizer, in which case $p_z$ is maximum, equal to 1. This option however has a major inconvenient: it does not provide a conversion of the vortex core gyration to microwave emissions. Indeed the global magnetic configuration is invariant when the core rotates, as illustrated on Fig. 19a.

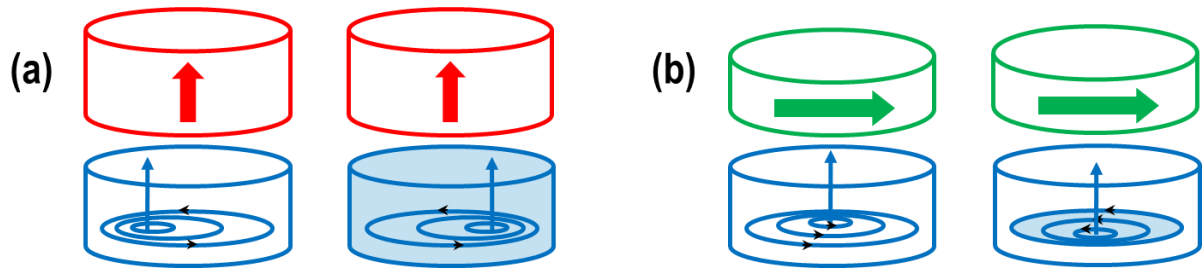

**Figure 19:** (a) A vortex precessing above a perpendicular polarizer does not induce resistance variations. (b) An in-plane sensing layer can translate the core gyration to resistive variations.

To solve this problem, a solution is to fabricate a hybrid excitation/detection structure. The vortex excitation is provided by the perpendicular polarizer/spacer/vortex trilayer. The vortex gyration detection is possible by adding a supplementary vortex/spacer/in plane magnetized sensing layer trilayer. As illustrated on Fig. 19b, the configuration sketched on the left is more parallel than the configuration sketched on the right, converting vortex gyration to resistance oscillations.

We have chosen to use a spin valve for vortex excitations and a magnetic tunnel junction for detection in order to benefit from the large resistance variations of the latter to convert vortex oscillations to a large microwave power (see Fig. 20 a and b). The result is shown on Fig. 20c. The integrated power, 0.6 µW, is larger than for the previous SAF / MgO / NiFe junctions thanks to the insertion of a thin CoFe layer between MgO and NiFe (TMR ~ 40 %). And, as predicted, we observe, at zero field, a highly coherent emission peak, of linewidth 590 kHz. We will see in the following section that the spectral purity of vortex oscillators can be further improved.

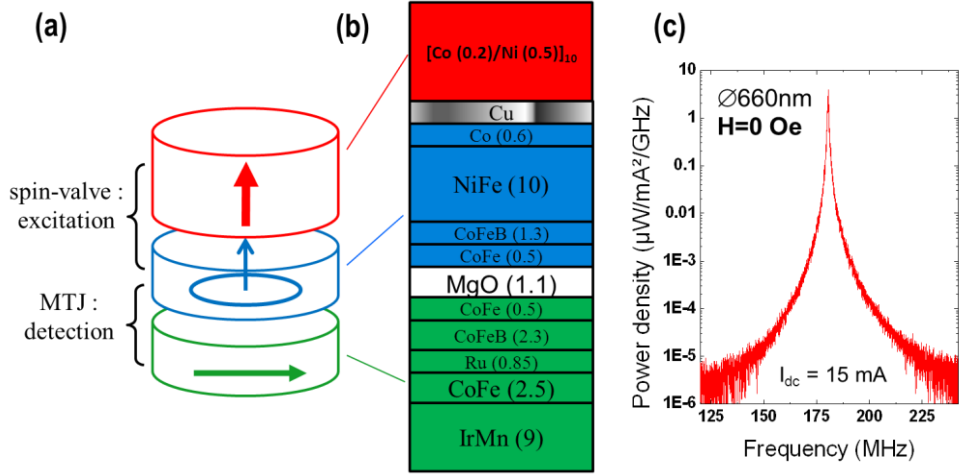

**Figure 20:** (a) Schematic of the hybrid structure for zero field vortex excitation. The top part is a spin valve designed to induce spin transfer vortex excitations in the free layer. The bottom part is a tunnel junction translating the vortex core oscillations to large resistance variations. (b) Sketch of the fabricated stack with the layer thicknesses in nm. (c) Microwave spectra at zero field and I = 15 mA.

> **A first solution to obtain spin torque induced microwave vortex oscillations at zero field is to use a perpendicular polarizer. In that case, the rotational symmetry of the junctions does not allow a conversion of the vortex rotation to magneto-resistance variations, and an additional sensing layer is needed. We have fabricated such SAF CoFeB sensing layer / MgO / NiFe vortex free layer / Cu / CoNi perpendicular polarizer, and obtained powerful (0.6 µW), narrow ( 590 kHz) emissions at zero field.**

### b. Non-uniform planar polarizer

The spin transfer force derived previously is no longer valid if the polarizer is not fixed or not uniform. Indeed, the energy dissipated or gained by spin transfer depends crucially on the magnetization landscape seen by each vortex spin when it flies over the polarizer during the rotation. We have transposed our method based on the dissipated energy calculation to the case of a spatially non-uniform planar polarizer [77].

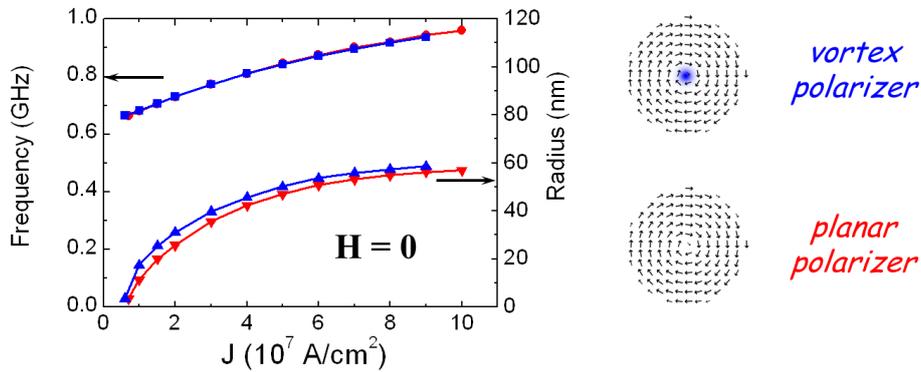

**Figure 21:** Micromagnetic simulations: frequency and radius corresponding to a vortex core precession excited by a non-uniform in-plane polarizer (blue : vortex polarizer, red : circular polarizer, equivalent of a « vortex without out-of-plane component »), at zero field, for a 200 nm diameter pillar.

The spin transfer force expression should be then re-written as: $\vec{F}_{ST}^{//} = \pi M_s L b P \sigma J [\vec{p}(\vec{X}) \cdot \vec{e}_\chi] \vec{e}_\chi$, with L the dot thickness, b the vortex core radius, P the vortex core polarity, and **p(X)** the in-plane spin polarization related to the non-uniform polarizer magnetization, and **e**$_\chi$ the tangential unit vector. When the polarizer is in-plane, only the vortex core, where the magnetization has a component out-of-plane, contributes to the spin transfer force. According to the above expression for the spin transfer force, it appears that $\vec{F}_{ST}^{//}$ will be maximized if **p(X)** is everywhere parallel to **e**$_\chi$, in other words if the polarizer itself is a vortex.

The micromagnetic simulations of Fig. 21 confirm the existence of zero field large amplitude vortex oscillations for a vortex polarizer. I have therefore fabricated, in collaboration with LPN Marcoussis, NiFe 4 nm / Cu 10 nm / NiFe 15 nm spin-valve nanopillars with 100 and 200 nm diameters. A schematic of the structure is presented on Fig. 22a. At zero current, none of the NiFe layer contains a vortex. The magnetizations are quasi-uniform and coupled anti-parallel by the dipolar field. When the current is ramped up, the Oersted field increases which favors vortex formation. The thick layer is the first to change its configuration, corresponding on Fig. 22a to a resistance jump to about half-magnetoresistance at $I_{dc} \sim 2$ mA. At larger current, the thin layer finally transforms to a vortex as well, and the resistance reaches a low value close to the uniform parallel state resistance. Fig. 22b shows the microwave spectra obtained for an in-plane field of 345 Oe and increasing currents. The impact of the transition for the uniform/one vortex configuration to the double vortex state is significant, with a strong increase of the spectral power density associated with a considerable decrease of the linewidth, that can dive below 50 kHz for certain values of field and current [78]. These observations highlight the interest of coupling two systems with low linewidths (here two vortices), to obtain an even larger spectral purity.

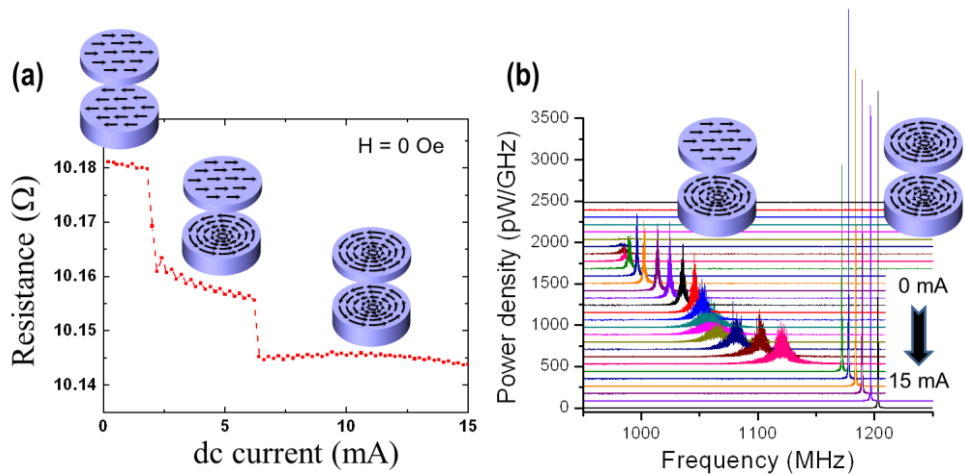

**Figure 22:** (a) Resistance versus injected current curve at zero field for a 100 nm diameter pillar. The evolution of the magnetic configuration with current is sketched. (b) Microwave spectra for an in-plane field of 345 Oe, and applied currents from 0 to 14.5 mA with a step of 0.5 mA.

---

**A second solution to obtain zero field spin torque induced vortex emissions is to use a non-uniform or an oscillating polarizer. We have derived an analytical expression for the spin torque force acting on a vortex in the case of a planar non-uniform static polarizer. It shows that the force is largest when the polarizer itself is a vortex. We have fabricated spin-valve nanopillars where, depending on the amplitude of the current-induced circular Oersted field, the magnetization configuration can be tuned from [uniform / uniform] to [uniform / vortex] to [vortex / vortex]. In the latter case we obtain large, well defined zero field oscillations. The associated low linewidth, that can dive below 50 kHz, demonstrates the interest of coupled systems (here the coupled oscillations of two vortices) for increasing the spectral purity.**

**Related publications:**

A.V. Khvalkovskiy, J. Grollier, N. Locatelli, Ya. V. Gorbunov, K. A. Zvezdin, and V. Cros, "Non-uniformity of a planar polarizer for spin-transfer-induced vortex oscillations at zero field", Appl. Phys. Lett. **96**, 212507 (2010)

N. Locatelli, V.V. Naletov, J. Grollier, G. de Loubens, V. Cros, C. Deranlot, C. Ulysse, G. Faini, O. Klein and A. Fert, "Dynamics of two coupled vortices in a spin valve nanopillar excited by spin transfer torque", Appl. Phys. Lett. **98**, 062501 (2011)

P. Skirdkov, A.D. Belanovsky, K.A. Zvezdin, A.K. Zvezdin, N. Locatelli, J. Grollier and V. Cros, "Influence of shape imperfection on dynamics of vortex spin-torque nano-oscillator", Spin **02**, 1250005 (2012) (2012)

# 5. Synchronization of vortex oscillators

### a. Phase locking of vortex nano-oscillators: experimental results

Thanks to their low linewidth, and correlated high emitted power when they are integrated in magnetic tunnel junctions, vortex oscillators seem ideal candidates for synchronization. We have therefore experimentally studied the phase locking properties of vortex oscillators to a microwave current. The samples are the previously discussed SAF/MgO/NiFe magnetic tunnel junctions with in-plane polarizer. As shown in Fig. 23a, while the emission frequencies of our vortex oscillators are close to 700 MHz, the observed locking ranges are of the order of 250 MHz for an injected microwave current of 800 µA. This means that vortex oscillators can be locked on a range representing the third of its frequency. This is much better than previous results obtained with any other system, and demonstrates a high coupling efficiency between the gyrotropic vortex mode and the microwave current.

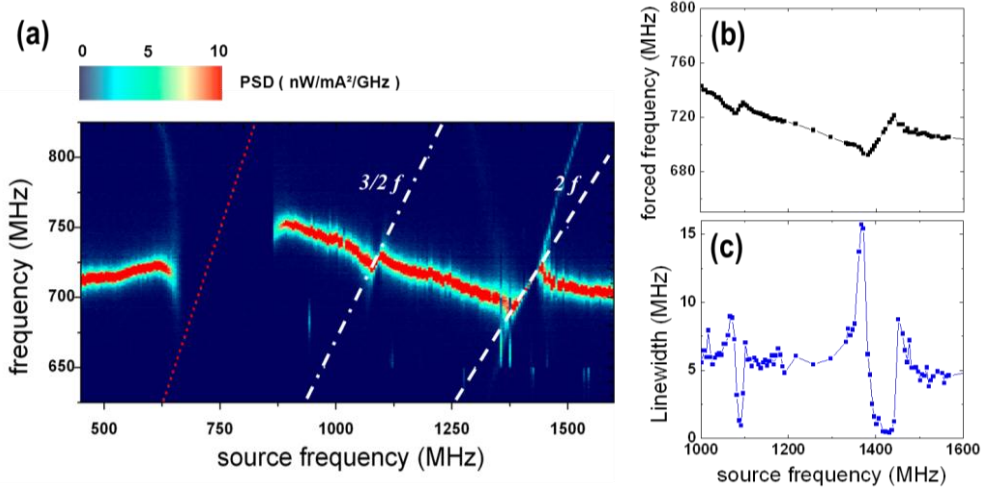

**Figure 23:** (a) Power spectral density variations as a fuction of the source frequency for $I_{dc}$ = 3.5 mA, $H_{perp}$ = 5.8 kOe and $I_{hf}$ = 0.8 mA. The dotted white lines indicate 3/2 $f_{source}$ and 2 $f_{source}$. (b) Evolution of the gyrotropic vortex core frequency as a function of the source frequency (c) Linewidth variations with the source frequency.

It appears on Fig. 23 that in addition to the standard synchronization for $f_{source}$ close to $f_{oscillator}$, fractional synchronizations occur for $f_{source}$ close to 3/2 $f_{oscillator}$ and 2 $f_{oscillator}$. These fractional synchronizations are interesting from a theoretical point of view, as an indicator of the respective symmetries of the mode and the excitation, but also from a practical point of view. Indeed, when $f_{source}$ is close to $f_{oscillator}$, the spectral peak of the oscillator is sucked into the peak of the source, which prevents its study. When fractional synchronization occurs, the oscillator peak is far from the source emission, and it is possible to characterize the influence of phase locking on the oscillator spectral properties. The evolution of the frequency and linewidth of the vortex oscillator as a function of the source frequency in the vicinity of 3/2 $f_{oscillator}$ and 2 $f_{oscillator}$ is shown on Fig. 23b and c. The linewidth drop in the locking zones is striking. By adjusting the spectrum analyzer resolution, we have been able to determine that the oscillator linewidth goes down to 3 kHz. This corresponds to a factor 1000 reduction compared to the free running oscillator, and a perfect synchronization during $2 \cdot 10^5$ periods.

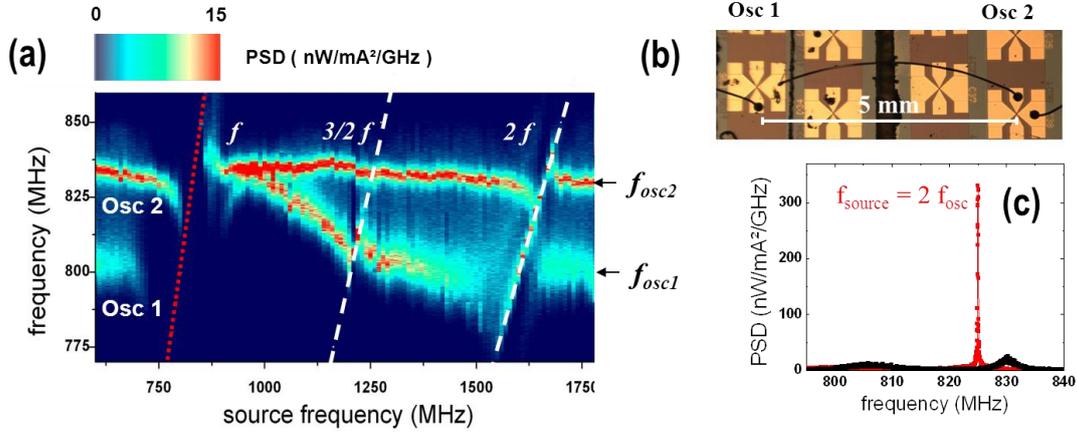

**Figure 24:** (a) Power spectral density emitted by two oscillators connected in series as a function of the source frequency for $I_{dc}$ = 3.5 mA, $H_{perp}$ = 5.8 kOe and $I_{hf}$ = 0.67 mA. The dotted lines correspond to 3/2 $f_{source}$ and 2 $f_{source}$. (b) Photograph of the two oscillators connected in series by gold wires. (c) Power spectra measured for $I_{dc}$ = 3.5 mA, $H_{perp}$ = 5.8 kOe and $f_{source}$ = 1650 MHz (red squares) and without external source (black squares).

The strong coupling to the source and the weak frequency dispersion of these oscillators based vortex core precessions have allowed us demonstrating the coherent oscillation of two oscillators simultaneously synchronized to a microwave source. In that experiment, two oscillators, several millimeters apart, are electrically connected in series by wire bonding (Fig. 24b). Fig. 24a shows the evolution of spectra as a function of the source frequency. Out of the synchronization zones (once again at f, 3/2 f, and 2f), the peak of each oscillator can be easily identified. At f and 2f, the two oscillators are simultaneously locked to the source. The measurement of the power spectra (Fig. 24c) with and without the microwave source shows the transformation of two small, broad peaks (black) to a thin single peak with high spectral power density (red). This experiment illustrates well the interest of synchronization for improving the spectral purity of spin transfer oscillators [79].

> **Phase locking experiments of our [SAF fixed polarizer] / MgO / [NiFe vortex] oscillators to a microwave source reveal that vortex oscillations couple extremely well to the external alternating current forcing, giving rise to wide locking ranges, reaching 1/3 of the oscillator frequency.**

**Related publications:**

### b. Non-linear large amplitude vortex gyration: model

Explaining the high efficiency of phase locking to a microwave current first requires having a comprehensive description of large amplitude vortex core oscillations. In particular, the coupling efficiency calculation is done by studying the phase perturbations on the oscillator limit cycle due to the external excitation. Before the start of spin transfer induced vortex precession studies, nobody spent time to describe accurately these large orbit trajectories. Indeed, all experimental studies were done at resonance, and therefore with relatively weak precession amplitudes.

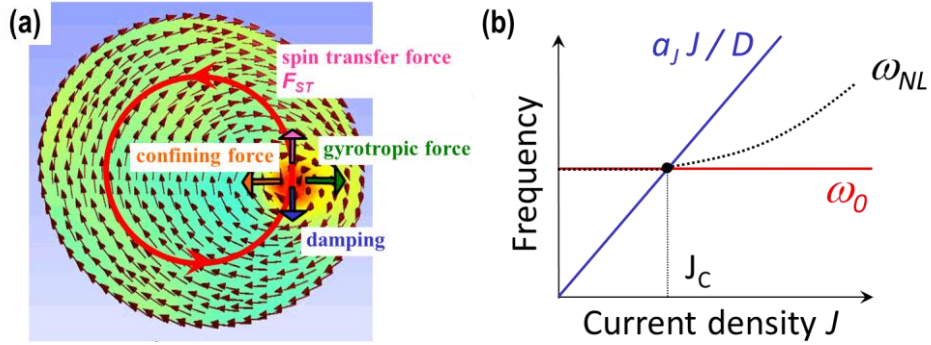

**Figure 25:** (a) Forces on the vortex core: sketch. (b) Frequency as a function of current. In blue and red: frequency variations when non-linearities are not taken into account (blue: balance of dissipative forces, red: balance of conservative forces). In black: schematic of a typical f(J) evolution when non-linearities are taken into account.

The balance of forces acting on the vortex core is reminded on Fig. 25. In a first step let's consider that only the linear terms in s = r/R (ratio of the orbit radius over the dot radius) are taken into account in the expression of the different forces. In that case, the radial balance of forces will impose a first condition on the frequency value: $\omega = \frac{\kappa}{G} = \omega_0$, where κ is the restoring constant arising from confinement, and G the gyrovector. This is the red line on Fig. 25b. Establishing the balance of forces in the direction tangent to the core trajectory will give yet another condition on the oscillator frequency: $\omega = \frac{a_J J}{D}$, where $a_J$ is the spin transfer efficiency, and D the vortex damping[1]. This condition is illustrated by the blue line on Fig. 25b. Clearly, these two conditions (blue and red) cannot be simultaneously fulfilled, except for one value of the current density: the critical current density, for which the orbit radius is zero. The linear vortex core dynamics does not allow describing correctly the motion as soon as the orbit radius is non-zero. If the non-linearities are now integrated into the expression of forces, the restoring constant becomes $\kappa(1 + \varsigma s^2)$ and the damping $D(1 + \xi s^2)$ (neither the gyrovector nor the spin transfer force contain non-linear terms). Then, the two previous conditions can be rewritten as: $\omega = \frac{\kappa}{G}(1 + \varsigma s^2)$ and $\omega = \frac{a_J J}{D(1+\xi s^2)}$. This corresponds for example to increasing a bit the frequency of the first condition, and decreasing slightly the frequency of the second, by adjusting the orbit radius so that the two frequencies finally coincide for all current densities (black line on Fig. 25b). This means that non-linearities are not only going to modify the vortex core frequency, but that they are also going to set the rotation amplitude. And yet, very few attempts to determine the exact values of the non-linearity coefficients have been performed, most of them purely analytical and not often compared to simulations or experiments, focused on one coefficient in particular and systematically disagreeing between each other [76, 80, 75, 81].

We have therefore decided to determine the non-linearity coefficients relative to vortex core precessions, as a function of the applied magnetic field, since this latter one is necessary to obtain oscillations in systems with fixed and uniform planar polarizer such as our

---

[1] The particular geometry of this magnetic object enhances its damping compared to the Gilbert value α.

SAF/MgO/NiFe samples. We chose to systematically compare analytical results, the values we extract from micromagnetic simulations and our experimental results. Some analytical expressions had already been calculated by other authors, and we simply validated them with micromagnetic simulations. This is the case for example of the linear confinement due to magnetostatic charges, obtained by Guslienko et al. [82] and the corresponding non-linear coefficient, calculated by Gaididei et al. [81], as well as the evolution of these parameters with an out-of-plane magnetic field, given in De Loubens et al. [83]. We had already calculated in [74] the linear coefficients corresponding to damping and Oersted field confinement. We have extended these calculations to obtain their non-linear counterparts plus their variations with an out-of-plane field. I will not here go into the details that can be found in [84], but we obtain a very good agreement between analytical calculations and micromagnetic calculations. Only the analytical calculation for the non-linear damping coefficient is not good enough to approach the value derived from micromagnetic simulations, probably because the two vortex ansatz does not provide a good description of large amplitude vortex deformations. We then inject the values extracted from micromagnetic simulations into Thiele equation, and calculate the evolution of orbit and frequency as a function of current, as well as the power variations as a function of the out-of-plane applied magnetic field. As can be seen on Fig. 26, there is an excellent agreement between experimental results, micromagnetic simulations and analytical model. Our results show that it is now possible to fully account for the non-linear dynamics of a vortex core.

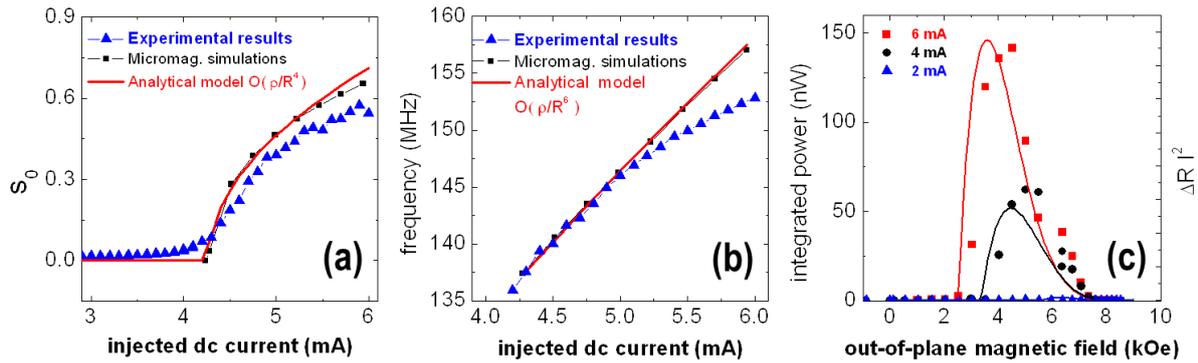

**Figure 26:** Evolution of the orbit radius (a) and frequency (b) as a function of the injected dc current. The experimental results are in blue, the micromagnetic simulations in black, and the analytical model in red. (c) Integrated power as a function of the out-of-plane applied magnetic field, for different current values. The symbols are experimental results, and lines analytical calculations.

This result will now allow us, by fully describing the limit cycle thanks to our analytical model, to include the phase perturbations induced by an external microwave current, in order to obtain a theoretical framework for vortex oscillators' synchronization.

> **To understand the excellent coupling of vortex oscillators to an injected microwave current, it is first necessary to be able to model correctly the large orbit spin torque induced vortex gyrations. This requires the knowledge of all non-linear coefficients governing the dependence of frequency and damping with orbit under an applied perpendicular magnetic field. We have analytically calculated the coefficients that were missing in the literature. We have checked by micromagnetic simulations the validity of our calculated coefficients as well as those obtained from other works. We have finally compared our calculations to our experiments. The very good agreement between analytical calculations, micromagnetic simulations and experiments shows that we are now able to describe correctly large spin torque induced vortex precessions.**

**Related publications:**

A. Dussaux, A. V. Khvalkovskiy, P. Bortolotti, J. Grollier, V. Cros and A. Fert, "Field dependence of spin-transfer-induced vortex dynamics in the nonlinear regime", Phys. Rev. B **86**, 014402 (2012)

### c. Synchronization of vortex nano-oscillators

Our first locking experiments, performed in 2008 and described in section II.2.c, showed the impact of linewidth and tunability on spin transfer nano-oscillators synchronization. These measurements on all-metallic spin-valves also indicated that the coupling between the quasi-uniform mode under study and the external microwave current was not very strong. Later measurements by S. Urazhdin et al. performed in very similar systems show that on the contrary, the quasi-uniform mode couples very well to the excitation provided by a microwave field, giving rise to wide locking ranges [85]. The respective symmetries of the oscillator vibration mode and the microwave excitation therefore play an important role in the coupling and phase locking efficiencies. The group of Olivier Klein and Gregoire de Loubens in CEA Spec has elegantly demonstrated and analyzed the role of these symmetries by measuring the local ferromagnetic resonance (MRFM experiments) of specially designed spin-valve pillars that we fabricated for that purpose [86]. The final structure includes an antenna above the pillar, giving the possibility to generate a microwave field in addition to simply injecting a microwave current through the pillar.

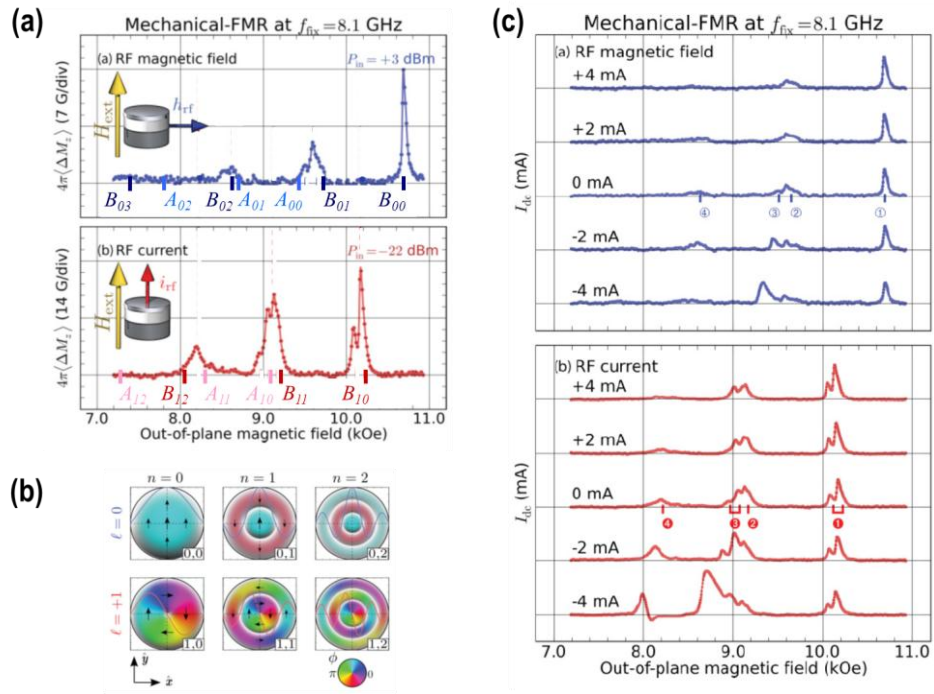

**Figure 27:** (a) Microwave spectra obtained by MRFM. The excitation is delivered by the spatially uniform microwave field generated by an antenna (upper panel), or by a microwave current injected in the pillar at the same frequency (bottom panel). The position of calculated modes is given by the azimuthal and radial indices (l,n). The letter A refers to the thin layer, B to the thick layer. (b) 2D profiles of the analytically calculated modes for different values of (l,n). The arrows represent the transverse magnetization at a given time t. The color codes for the direction of this transverse magnetization. (c) Same as a, in the presence of an injected dc current.

Fig. 27a compares the spectra obtained by the two methods, under an applied out-of-plane static magnetic field. It appears immediately that each excitation leads to the vibration of very different modes. Indeed, the spatially uniform microwave field can only couple to modes with an azimuthal index l = 0 (see mode profiles on Fig. 27b). For the large experimentally applied fields, the two magnetizations are saturated out-of-plane. Since they are parallel, the microwave current cannot generate a spin transfer torque, but it creates a circular Oersted field. The latter, by symmetry, can only couple to modes with an azimuthal index of 1. These symmetry considerations are confirmed by analytical calculations and 3D micromagnetic simulations giving the field value corresponding to each mode at the given frequency of excitation. The modes, which profiles are shown on Fig. 27b, are classified by their azimuthal and radial index. The letter A refers to the thin layer, B to the thick layer. It can be seen on Fig. 27c that injecting a dc current allows discriminating the thin and thick layer modes. Indeed, with the convention adopted here, by spin transfer, a positive current will destabilize the thick layer, and a negative current the thin layer. Fig. 27c shows that the amplitude of the peaks labeled B, corresponding to the thick layer, increases for positive currents and decreases for negatively currents, and inversely for the thin layer A peaks.

An important conclusion of this study concerns synchronization. Let's take the example of our earlier 2008 work on phase locking by an injected microwave current in spin valves [59]. In that study, we apply an out-of-plane field, not too strong in order to keep an angle between the two magnetizations, and obtain thanks to the dc spin transfer torque an efficient sustained vibration of the lower energy (0,0) uniform mode. When a microwave current is injected, it generates an AC spin transfer torque, plus a circular Oersted field. The AC spin transfer torque, spatially uniform, has the good symmetry to couple to the (0,0) mode : it is at the source of the locking phenomena we had observed experimentally. Unfortunately, the Oersted field that can be even larger (especially for small angles between magnetizations), does not back up the microwave spin torque since it can only couple to l = 1 modes. Therefore the coupling is not optimal between the mode excited by dc spin transfer, and the modes to which can couple the microwave injected currents. That's why our measured locking ranges are much smaller than the ones observed by Urazdhin et al [85]. Indeed, in this latter study, the applied microwave field couples perfectly to the (0,0) mode excited by the dc spin transfer torque. This means that to obtain an efficient synchronization, it is important to design the system so that the dc spin transfer torque induces a vibration mode to which the microwave excitation can couple efficiently. Another way to say it is that it's not because the microwave excitation is large that it will create a large force on the vibrating mode. But we are going to see that fulfilling this condition is not enough. Even if the force generated by the external source on the oscillating mode is large, there is another major condition to obtain synchronization. We will see in the following the importance of these considerations for the phase locking of vortex oscillators.

At the origin of synchronization is the modification of an oscillator's phase by an external AC excitation. Synchronization is therefore only possible if the force generated by the external AC excitation depends both on the phase of the external signal $\omega_s t$ AND the phase of the oscillator θ. We are studying circular dots containing a vortex. Clearly, the microwave Oersted field has a good symmetry to couple to the vortex gyration mode, and indeed it generates a large force on the vortex core. But the action of the microwave Oersted field does not depend on the vortex core position. Depending on its own phase $\omega_s t$ only, it will tend to confine or eject the core, independently of the oscillator's phase. The Oersted field cannot contribute directly to phase-lock vortex oscillators. In our samples with a uniform SAF polarizer, the spin transfer force that induces the vortex sustained precession is provided by the Slonczewski torque, through the perpendicular component of spin polarization:
$\vec{F}_{ST}^{\perp} = j p_z \sigma \pi M_s (\vec{e}_z \times \vec{X})$. The amplitude of this force is also independent on the oscillator phase, and cannot be at the origin of the large locking ranges observed experimentally. Two other forces can contribute to synchronization. The first one is the spin transfer force induced by the in-plane torque through the action of the in-plane spin polarization. When a dc current is applied, the latter has no net effect, because its sign is opposed for each half-trajectory. But if the current now becomes AC, this cancellation effect will be prevented, and the resulting force will depend on the respective phases of the oscillator and microwave current: $\vec{F}_{ST}^{//} = \Lambda_{ST}^{//} cos(\omega_s t) sin\theta \vec{u}_\theta$, where $\Lambda_{ST}^{//} = -ln2\pi\, CM_s LJ_{ac} bPp_x$ according to the expression of the force due to an in-plane polarizer that we calculated in [77] and mentioned in section II.4.b. (C : vortex chirality, $M_s$ saturation magnetization, L free layer thickness, $J_{ac}$ microwave current, b vortex core radius, P core polarity and $p_x$ in-plane spin polarization). The second force that can contribute to synchronization originates from the second spin transfer torque: the out-of-plane field-like torque. When a dc current is injected, the component of this field like torque in the plane of the layers will slightly shift the vortex core away from the dot center. In magnetic tunnel junctions with asymmetric electrodes, such are the ones under study (CoFeB/MgO/NiFe), the field-like torque can depend linearly on current, and even reach large values [87] [21]. So if an AC current is injected, this force will act like an effective microwave field along the polarizer. Its in-plane component will break the axial symmetry, so that its action on the core will depend on the oscillator phase: $\vec{F}_{FLT}^{//} = \Lambda_{FLT}^{//} cos(\omega_s t)(cos\theta\, \vec{u}_\theta - sin\theta\, \vec{u}_\rho)$, with $\Lambda_{FLT}^{//} = -\frac{2}{3}\pi\, CM_s LRJ_{ac}\lambda p_x$ ($\lambda$ is the ratio between the field like torque and the in-plane torque). The ratio between the two forces that can be at the origin of synchronization: $\frac{\Lambda_{FLT}^{//}}{\Lambda_{ST}^{//}} = \frac{R\lambda}{b}$ indicate that in an asymmetric magnetic tunnel junction, the effective field can be most efficient, since the dot radius R is large compared to the core radius b, and that in addition $\lambda$ is not negligible.

Following the same method as we did for the uniform mode (see section II.2.c and [59]), we obtain the Adler phase equation from the non-linear vortex dynamics: $\frac{d(\Delta\varphi)}{dt} = 2\pi(f_0 - f_{source}) + \varepsilon\, cos(\Delta\varphi)$, where $\Delta\varphi = \varphi_{osc} - \varphi_{source}$ and $\varepsilon$ represents the coupling between the oscillator and the source. The locking range is then equal to 2 $\varepsilon$, and we determine that: $\varepsilon = \frac{1}{s_0 RD}\frac{\varsigma}{\varsigma+\xi}\sqrt{\left(\Lambda_{ST}^{//}\right)^2 + \left(\Lambda_{FLT}^{//}\right)^2}$.

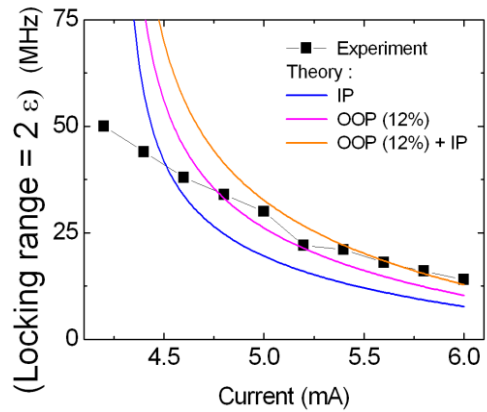

**Figure 28:** Locking range as a function of the dc current for $H_{perp}$ = 3.2 kOe. Experimental results (black squares). Calculation taking into account the in-plane torque IP (blue curve), the field like torque OOP (pink curve), and both contributions (orange curve).

Fig. 28 shows the comparison between experimental results and the theoretical prediction for the locking range. Above the critical current (~ 4.2 mA), the calculated expression for 2ε gives an excellent approximation of the experimental results if both spin transfer torques (IP and OOP) are taken into account, and the ratio λ is set to 0.12 (orange curve). This latter value for λ is in good agreement with values derived from spin diode experiments performed on magnetic tunnel junctions with slightly different layer composition [88].

In addition, in the case of a phase locking originating purely from the effective field term, the predicted phase difference between the oscillator and the source is zero [89], which is extremely favorable for the self-synchronization of an assembly of magnetic tunnel junction oscillators connected in series [90].

> **We have seen that conditions to obtain an efficient phase locking of spin torque oscillators to an injected microwave current are a low linewidth and a good tunability. The fact that vortex oscillators possess both qualities does not entirely explain their excellent coupling compared to the modest results obtained for the uniform mode. In fact, low phase noise and large non-linearity are required but not sufficient for synchronization, and there are two more conditions. The first one is to adjust the symmetry of the external forcing to the symmetry of the vibrating mode. In other words, even if the amplitude of the external source is very large, it can have very little action on the vibrating mode. The second condition is that the resulting external force on magnetization has to depend on the phase of the oscillator. Otherwise, at a given time t, whatever the excursion of the oscillator on its limit cycle, the external force will have the same action, and will not be able to tune the oscillator phase. We have used our full description of the vortex oscillations to extract the associated phase dynamics under microwave current injection. We found that vortices can efficiently phase lock to this alternating external current thanks to the complementary actions of in-plane and out-of-plane spin torques.**

**Related publications:**

V.V. Naletov, G. de Loubens, G. Albuquerque, S. Borlenghi, V. Cros, G. Faini, J. Grollier, H. Hurdequint, N. Locatelli, B. Pigeau, A.N. Slavin, V.S. Tiberkevich, C. Ulysse, T. Valet and O. Klein, "Identification and selection rules of the spin-wave eigenmodes in a normally magnetized nanopillar", Phys. Rev. B **84**, 224423 (2011)

A. Hamadeh, G. de Loubens, V.V. Naletov, J. Grollier, C. Ulysse, V. Cros, O. Klein, "Autonomous and forced dynamics in a spin-transfer nano-oscillator: Quantitative magnetic-resonance force microscopy", Phys. Rev. B **85**, 140498(R) (2012)

## 6. Conclusion on spin-transfer nano-oscillators

By applying our condition: $\frac{\Delta R_{osc}}{R} > \frac{2\,(LW+D)}{I_{dc}} \frac{1}{\left(\varepsilon/I_{hf}\right)}$ for synchronization of serially connected oscillators to our SAF/MgO/NiFe vortex oscillators, using experimentally derived parameters, we find that we are just reaching the condition limit, and that electrically synchronizing these oscillators is now at reach. Indeed, $P_{int}$ = 5 nW corresponds to $\Delta R_{osc}/R$ ~ 0.33 %, for our 50 Ω oscillators [41]. With a linewidth of 1 MHz, a dispersion of 1 MHz, a dc current of 3 mA, and a coupling efficiency ($\varepsilon/I_{hf}$) = 500 MHz/mA [79], the right hand side of the equation is equal to 0.26 %. Improving slightly the microwave characteristics of vortex oscillators (by using the hybrid GMR/TMR strcutures for example) will facilitate synchronization. In particular, increasing the TMR ($P_{int}$ = 5 nW corresponds to samples with 15 % TMR) and reducing the critical currents for oscillations will be crucial. Indeed, it should be stressed that the loss of TMR at high bias ($\Delta R/R$ = 15 % at low bias is reduced to an oscillating ratio $\Delta R_{osc}/R$ of merely 0.33 % at large bias) makes synchronization complicated for large currents.

As we will see in the research project section, synchronization of several oscillators is an important step to achieve, not only for their applications as nano-scale microwave sources, but also for their use as computational blocks in associative memory processors.

## II. MEMRISTORS

### 1. From Spin-Torque nano-oscillators to memristors

As we have seen in the first part, spin torque allows building nano-devices with a wide range of operations: binary memory, stochastic device, microwave oscillator and spin wave emitter. Spin torque can also be used for microwave detection. Indeed, as illustrated in Fig. 29, it suffices to replace the injected dc current by an injected microwave current to achieve signal frequency detection. This reverse effect of spin transfer emission called spin torque diode has been demonstrated in 2005 [91]. Easily measurable dc voltage amplitudes of several hundred microvolts have been reported [92]. The conversion efficiency of the injected microwave power into dc voltage can be over 500mV/mW, outperforming semiconductor Schottky diodes.

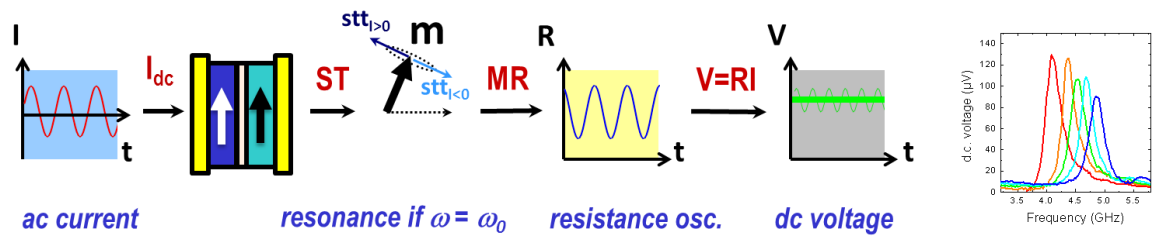

**Figure 29:** Principle of microwave detection by spin transfer torque (spin diode effect). If the frequency of the injected microwave current closely matches the eigenfrequency of the free layer vibration mode, the induced magnetization motion can be strongly amplified through resonance. Through this process, the alternating injected current induces resistance oscillations at the same frequency, leading to the appearance of a dc rectified voltage signal. Right: Experimental curve showing the dc voltage as a function of frequency. For different applied fields, corresponding to the different colored curves, the resonance frequency varies, resulting in a shift of the peak.

This comes to show that after more than a decade of intense research, the understanding of spin torque's microscopic origins and the resulting magnetization dynamics has reached such a level of maturity that it is now possible to predict accurately through coupled transport/micromagnetic simulations the device behavior. Smart engineering of materials, sample geometry and device's response can provide new bricks for implementing novel spin torque devices functionalities. Tuning different parameters such as input current waveform, materials (spin valve or magnetic tunnel junction, in-plane or out-of-plane magnetized layers etc.), geometry (pillar, point contact, stripe) allows to design specific functions at the nano-scale.

In this context, in 2008, a lot of buzz surrounded a publication by a Hewlett-Packard team in Palo Alto [93]. The authors claimed to have fabricated a new kind of nano-device, the missing "memristor", and hinted at the capacity of this component to mimic biological synapses. Soon after, the huge DARPA SyNAPSE program was engaged, pointing out the crucial importance of emulating synapses at the nanoscale for being able at last to build high performance brain-inspired hardware.

It seemed to me interesting to take advantage of the possibilities offered by Spin Torque to implement a magnetic version of a memristor, with all the advantages of spintronics: speed, purely electronic commutation, non-volatility etc. At the same time, colleagues from the Oxide group in the lab were obtaining their first results of Giant Tunneling Electro-Resistance. As everybody liked the idea of building nano-synapses, we soon agreed to try to exploit the several orders of magnitude resistance variations obtained in ferroelectric tunnel junctions when the electric polarization is flipped to develop a ferroelectric memristor.

Before going into the details of the two purely electronic memristor concepts and our experimental results, I will briefly describe why building brain-inspired hardware is interesting and relevant and how memristor nano-devices could be disruptive in this context.

## 2. Context: the re-birth of Artificial Neural Networks

Imitation using artificial tools of some of the incredible human brain capacities for cognitive processes will certainly be the next big challenge in the revolution of information technology. A few days-old baby is able to interpret and analyze sequences of images in a fraction of second whereas such apparently simple tasks are still out of the reach of the most powerful digital computers. In order to solve some problems of artificial intelligence, complex computational models called Artificial Neural Networks (ANNs) have been proposed based on the brain architecture. Even if ANNs with multiple neural layers are extremely powerful, they lose the speed, fault tolerance and low power consumption provided by their originally analog and parallel architecture when they are implemented as software simulations on our sequential computers. So as soon as the 80's, many groups started to fabricate hardware versions of artificial neural networks, a well-known example being the Intel ETANN chip [94]. But these first results were a huge disappointment. The neural networks models that were implemented at that time, perceptrons, had very limited performances. And, above all, these chips, developed during the boom of general processors, could really not compete against desktop computers.

Today, the context has changed. It is now well known that the CMOS industry is facing a number of hurdles to continue improving processors in terms of size, performance and dissipation. While the number of transistors keeps increasing, since 2005, their frequency is stalling due to the extremely large thermal dissipation. In order to keep on increasing performances, computer architects have introduced parallelism in their systems via multi-core architectures. Nevertheless other technical complications will have to be solved such as dark silicon [95] (almost 25% of the chip must be turned off at 22nm and by 8nm this could affect almost half of it) and the increasing number of defective components due to their shrinking dimensions. Still, with the information explosion, the need for less power consuming yet computationally strong architectures is at its peak. So developing large scale hardware neuromorphic architectures, for fast, low power, defect tolerant computing is now very relevant again. The need to develop new hardware devices with a brain-inspired massively parallel, dynamical architecture and radically different from contemporary IT technology is recognized by different communities, such as computer scientists, neuro-scientists, hardware designers. ANNs are in the short term extremely promising as task-specific accelerators to be integrated in conventional multi-core architectures. Due to the increasing amount of digital data, Intel has recently pointed out the importance of RMS (Recognition, Mining and Synthesis) applications [96]. There is a critical need for scalable, adaptable and programmable computing architectures that have the capability to model, classify and synthesize complex digital data, and recent Neural Networks algorithms such as "deep networks" [97], are intrinsically adapted to these kinds of tasks [98].

Neural networks are based on neurons, and synapses. Neurons are processing units that integrate information sent from other neurons through synapses, and spike when a threshold reached. Synapses transmit information according to their synaptic weight. The synapses are plastic, and it is the process of adjusting the weights to converge to a desired output for a set of given inputs that allows learning. The network memory is defined by the ensemble of synaptic weights. The performance of a Neural Network will depend on its size and interconnectivity. For example there are about $10^{11}$ neurons in the brain, and close to $10^4$ synapses per neuron. The difficulty to provide large numbers of dynamical synaptic connections has slowed down the development of neuromorphic circuits, which are still far from the performances of their biological counterparts. Today's neuromorphic chips are entirely based on CMOS technology. In particular, implementing the synaptic plasticity requires several tens of transistors [99]. The recent hardware ANN designs remain mostly focused on time-multiplexed implementations with multiple neurons mapped to a few hardware ones, and synaptic weights stored in SRAM banks. Using dynamical analog, reconfigurable nanoscale devices for the synaptic nodes would result in tremendous gains in terms of power, dissipation, miniaturization and computational efficiency. In 2008, the Hewlett-Packard team of S. Williams has demonstrated such devices, called "memristors" [93] [100]. Memristors could be the key to future developments of neuromorphic circuits.

> **Due to their application scope extremely relevant in the present context of data explosion: data recognition and mining, and thanks to their inherent qualities: speed, low power consumption and defect tolerance, there is a revival of neuromorphic hardware architectures. The performances of these massively parallel artificial neural networks on chip depend on their size and interconnection degree. It is therefore very important to have small elements to emulate the plastic synaptic interconnections. Memristor devices, that can mimic synapses at the nano-scale, can be the key to the future implementation of hardware neural networks as accelerators of next generation processors.**

### 3. Purely electronic memristors

A memristor is a tunable analog non-volatile resistance such that v=M(q)i [101]. The more intense the current through the structure, and the longer it is applied, the larger the resistance variations. This specific behavior explains the name "memristor" (memory resistor) and allows for example implementing the fact that the more a synapse is used, the better it transmits information (case of an excitatory synapse). In addition memristive behaviors are typically observed at the nanoscale (the HP devices are ~ 30x30 nm$^2$), which is crucial as we have just seen. The expression v = M(q) i is conferring memristors very particular transport curves, such as illustrated on Fig. 30. The i-v curves are "pinched" and the loop area is varying when the sweeping maximum current or frequency is varied (Fig. 30a). Another presentation of the same phenomena is given on Fig. 30b, where the resulting multi-state hysteretic resistance versus voltage curves are shown.

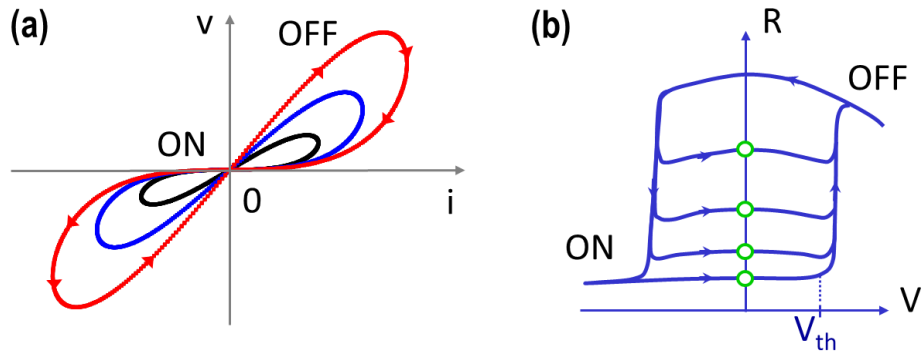

**Figure 30:** Characteristic transport features of memristors (a) pinched iv loops (calculated for a linear memristance M = a q + b) for different values of the maximum injected current (b) multi-state resistance versus voltage cycles.

The Hewlett-Packard work has motivated several groups to develop their own memristors for computing purposes [102] [103] [104] [105] [106] [107] [108] [109] [110]. All these devices belong to the same class of memories called Resistive Random Access Memories (RAMs). As most Resistive RAMs, almost all existing memristor concepts are based on physical phenomena that involve defects or deep material structural changes to induce the resistance variations: fuse/antifuse, nano-ionic or thermal processes (RedOx RAMs, Phase Change memories). For example, the memristive effects observed in Hewlett-Packard devices were first attributed to oxygen ions electromigration in an initially insulating $TiO_2$ matrix [100]. Using these devices will certainly require solving reliability issues related to large local heating, imprecise resistance control, component degradation and low write speed. While their performances may be sufficient for certain class of memories, the endurance and cyclability of ReRAMs may not be large enough for applications where the devices need to be written and rewritten repetitively. This is the case for cache memories, and for memristor synapses in unsupervised on-line neural networks were learning and adaptation is constant. There is therefore a practical interest to develop novel types of memristors, based on different physical concepts.

That's why I have started to develop "purely electronic" memristors, in other words nano-devices where the resistance changes are obtained through electron mediated phenomena at interfaces. These memristors promise an increased endurance and reliability, since the material structure is preserved, but also a faster commutation speed. In 2009, we have patented two new concepts: the "spin torque" and "ferroelectric" memristors [111] [112]. Both are based on emerging digital memory concepts, subject of intense academic and industrial developments. The "spin torque" memristor is derived from the Spin Torque MRAM, that should be on the market in 2013, and is foreseen as DRAM's replacement as cache memory [113]. As we have seen previously, the building block of ST-MRAM, the magnetic tunnel junction, relies on magnetization switching. The "ferroelectric" memristor is based on the ferroelectric resistive RAM [114]. Here the brick is the ferroelectric tunnel junction: an insulating ferroelectric ultrathin barrier sandwiched between two metallic electrodes. The commutation of polarization in the barrier when a voltage is applied across the junction can give rise to large resistance changes [115]. The idea is to transform these binary memories in multi-state, quasi-analog memristors. For that purpose, we will play with the mechanisms of ferromagnetic and ferroelectric commutation. We will design the devices in such way that switching occurs through non-uniform magnetic or ferroelectric domain configurations. The device resistance will then be directly related to the details of this domain configuration. Engineering and controlling these memristors will require understanding and controlling the magnetization/polarization dynamics in these systems.

**Building purely electronic memristors in contrast to the existing resistive switching technologies should upgrade their performances in terms of endurance and speed. We have proposed two memristors where the resistance variations are due to purely electronic effects: the spin torque memristor, based on the magnetic tunnel junction and the ferroelectric memristor, based on the ferroelectric tunnel junction. In both cases obtaining the multi-state, quasi-analog resistance variation of memristors will be achieved through a fine control of the nanoscale domain configuration during magnetization or polarization reversal.**

**Related publications:**

J. Grollier, V. Cros et F. Nguyen Van Dau, "Memristor device with resistance adjustable by moving a magnetic wall by spin transfer and use of said memristors in a neural network", French patent n°09 02122 (30/04/2009), international: WO 2010125181 A1,

M. Bibes, J. Grollier, A. Barthélémy et J-C. Mage, "Ferroelectric device with adjustable resistance", French patent n° 09 02845(2009), international: WO 2010142762 A1

## 4. Spintronic Memristor

### a. Principle of the spin torque memristor

A magnetic tunnel junction is a two state memristor, where resistance changes can be driven by dc current injection and subsequent spin torque effects (see Fig. 5a). In order to obtain the multistate analog features of memristors other groups have proposed to combine the bi-stable magnetic commutation with the analog resistive switching arising from redox / electromigration phenomena in the barrier [116] [117]. But in that case, the advantages due to "purely electronic" switching are lost. I have proposed, in order to obtain multi-resistance states while keeping the purely electronic writing, to use spin transfer induced domain wall motion. Let's consider a magneto-resistive trilayer with a domain wall in its free layer. The trilayer resistance depends on the relative proportion of parallel and anti-parallel domains, which is set by the domain wall position. As can be seen on Fig. 31a, the more the domain wall is to the left, the closer the configuration is to the parallel state, the smaller the resistance. By spin transfer effect, it is possible to manipulate the position of a magnetic DW [118] [119] [120] [121]. The domain wall displacement $\Delta x$ then depends on the amplitude j of the injected current as well as the pulse duration $\Delta t$: $\Delta x \propto j\Delta t = q$ [122] [123]. Therefore the resistance depends on the charge, and a spin torque memristor is obtained. The classical way to move a DW by spin transfer is to inject the current laterally, as illustrated on Fig. 31b. If the trilayer is a metallic spin-valve, the same lateral geometry allows reading the device resistance, thanks to Current-In-Plane (CIP) magneto-resistance. This is the trick we had used during my Ph.D. thesis in order to monitor the DW position, and show for the first time current-induced DW motion in a nanostructure [124] [125]. Some authors have proposed to use this geometry to implement the spin torque memristor [126] [127]. Unfortunately, the CIP magnetoresistance ratios in spin-valves are limited to a few %, which would give rise to OFF/ON ratios well below 1. This is much too low for discriminating the different states in a real-world application, and even more for implementing these devices in crossbar arrays. The solution to increase the OFF/ON ratio up to 6 and more is to use magnetic tunnel junctions [25] [26] [128]. In that case reading can only be achieved by applying the probe current vertically across the junction. The spin torque memristor concept that I proposed is based on vertical writing as well, as illustrated on Fig. 31c. This has two advantages. First, since the reading and writing paths are the same, we keep a two-terminal, easy to scale down, conform to definition memristor. Secondly, spin torque induced domain wall motion by vertical injection is much more efficient than the lateral scheme.

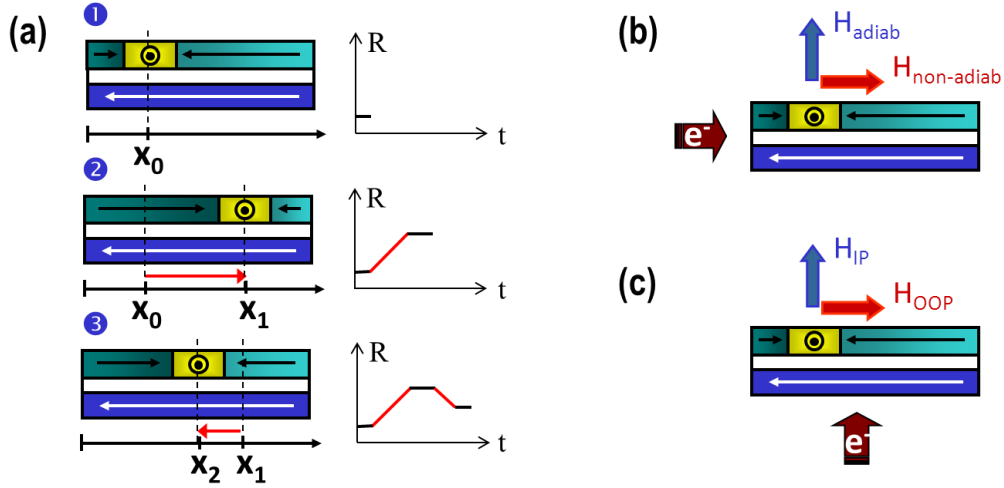

**Figure 31:** Spin Torque Memristor Principle (a) Schematic of the resistance variations linked to changes of the DW position (b) Schematic of the lateral current injection geometry and the effective fields related to the adiabatic and non-adiabatic spin torques (c) Schematic of the vertical current injection geometry and the effective fields related to the in-plane and out-of-plane spin torques.

**b. Lateral versus vertical spin injection for domain wall motion**

This is what we have pointed out in ref. [129]. A torque can always be related to an effective magnetic field through the relation: **T** = -γ **m** x **H**$_{eff}$, where m is the local magnetization. There are two conditions for a torque to move efficiently a DW: the torque should be orthogonal to the spins in the DW, and the related effective field should point along the domains magnetization. In the lateral injection configuration of Fig. 31b, there are two torques acting on the DW, called adiabatic and non-adiabatic torques for historical reasons. $\vec{T}_{adiab} = u\,\vec{m} \times \left(\vec{m} \times \frac{\partial \vec{m}}{\partial x}\right)$ and $\vec{T}_{non-adiab} = \beta u\,\vec{m} \times \frac{\partial \vec{m}}{\partial x}$, where the parameter u, linked to the amplitude of the spin polarized current, has the dimensions of a velocity: $u = JPg\mu_B/(2eM_s)$, with J the current density and P the spin polarization. As shown in Fig. 31b, the effective field related to **T**$_{adiab}$ is perpendicular to the domains. That is why, except for very particular cases [130], it cannot displace the DW efficiently [131]. On the contrary, the so-called "non-adiabatic" torque, **T**$_{non-adiab}$, has the proper symmetry for moving the domain wall: as illustrated on Figure 31b, its equivalent magnetic field points along the domains magnetization. Unfortunately, the parameter β has been evaluated experimentally [132] [133] [134] [135] [136] [137] [138] and theoretically [139] [140] [141] [142] [143] [144] [17] to be very low, typically close to the damping value α: β ~ 0.01-0.05. The efficiency of spin transfer to move the DW is consequently extremely limited in this lateral geometry: even if the spin polarized current is high, giving rise to large values of u, the torque acting on the DW is reduced by the small β factor, and the resulting spin torque βu is weak. The domain wall velocities, that can be written as v$_{DW}$(lateral) = βu/α, are experimentally limited to 150 m/s for current densities above $10^8$ A/cm$^2$, close to the electromigration threshold.

In the vertical current injection geometry, the two torques acting on magnetization have been already been introduced: they are the in-plane **T**$_{IP}$ and out-of-plane **T**$_{OOP}$ torques. A rapid analysis reveals that only the out-of-plane torque has the proper symmetry to move a DW in the case illustrated on Fig. 31c. In all metallic pillars, this out-of-plane torque is practically zero, and a vertical dc current injection is inefficient to move a DW [145] [146]. But in tunnel junctions, as mentioned previously, this torque can be large: $\eta$, the ratio between the in-plane and out-of-plane torques can reach more than 20 % in MgO based tunnel junctions [19] [20]. If we write the expression for the velocity in the vertical injection geometry, obtained from a 1D model, we find that: $v_{DW}$(vertical) = $\eta$ ($\Delta$/t) u/$\alpha$, where $\Delta$ is the domain wall width and t the free layer thickness. This expression immediately demonstrates the efficiency of vertical injection compared to lateral injection for domain wall motion. First, $\eta$, contrarily to $\beta$, is not small. Secondly, the geometrical factor $\Delta$/t amplifies the spin torque efficiency. Indeed the width of a transverse wall, typically over 100 nm, is much larger than the free layer thickness, typically below 5 nm. By picking reasonable numbers, (t = 5 nm, $\Delta$ = 50 nm, $\alpha$ = 0.01, $\beta$ = $\alpha$, $T_{OOP}$ = 0.1 $T_{IP}$), it can easily be seen that $v_{DW}$(vertical) ~ 100 $v_{DW}$(lateral) for identical current densities. Another way to emphasize the advantages of vertical injection is to write the critical current densities: $j_c$(vertical) ~ 0.01 $j_c$(lateral). We therefore expect two orders of magnitude gain in critical current densities in the vertical injection configuration.

> **The spin torque memristor is based on the current controlled displacement of a single magnetic domain wall in the free layer of a magnetic tunnel junction. A vertical current injection is used for reading and writing. Indeed we predict by simple symmetry arguments regarding which torque drives the domain wall that this geometry is more efficient to move a domain wall than the classical lateral current injection.**

**Related publications:**

A.V. Khvalkovskiy, K. A. Zvezdin, Ya.V. Gorbunov, V. Cros, J. Grollier, A. Fert, and A. K. Zvezdin,"High Domain Wall Velocities due to Spin Currents Perpendicular to the Plane", Phys. Rev. Lett. **102**, 067206 (2009)

S. Laribi, V. Cros, M. Muñoz, J. Grollier, A. Hamzic, C. Deranlot, A. Fert, E. Martínez, L. López-Díaz, L. Vila, G. Faini, S. Zoll and R. Fournel, "Reversible and irreversible current induced domain wall motion in CoFeB based spin valves stripes", Appl. Phys. Lett. **90**, 232505 (2007)

C. K. Lim, J. Grollier, T. Devolder, C. Chappert, V. Cros, A. Vaurès, A. Fert, G. Faini, "Domain wall displacement induced by sub-nano second pulsed current", Appl. Phys. Lett. **84**, 2820 (2004)

J. Grollier, P. Boulenc, V. Cros, A. Hamzic, A. Vaurès, A. Fert, G. Faini "Spin-transfer-induced domain wall motion in a spin valve", J. Appl. Phys. **95**, 6777 (2004)

J. Grollier, P. Boulenc, V. Cros, A. Hamzic, A. Vaurès, A. Fert, G. Faini, "Switching a spin-valve back and forth by current-induced domain wall motion", Appl. Phys. Lett. **83**, 509 (2003)

J. Grollier, D. Lacour, V. Cros, A. Hamzic, A. Vaurès, A. Fert, D. Adam, G. Faini, "Switching the magnetic configuration of a spin valve by current-induced domain wall motion", J. Appl. Phys. **92**, 4825 (2002)

### c. Bias dependence of spin transfer torques in magnetic tunnel junctions

For the purpose of fabricating a spin torque memristor, we want to be able to move the domain wall, via the OOP torque, in both directions. So we need a configuration where the OOP torque has a large linear bias dependence.

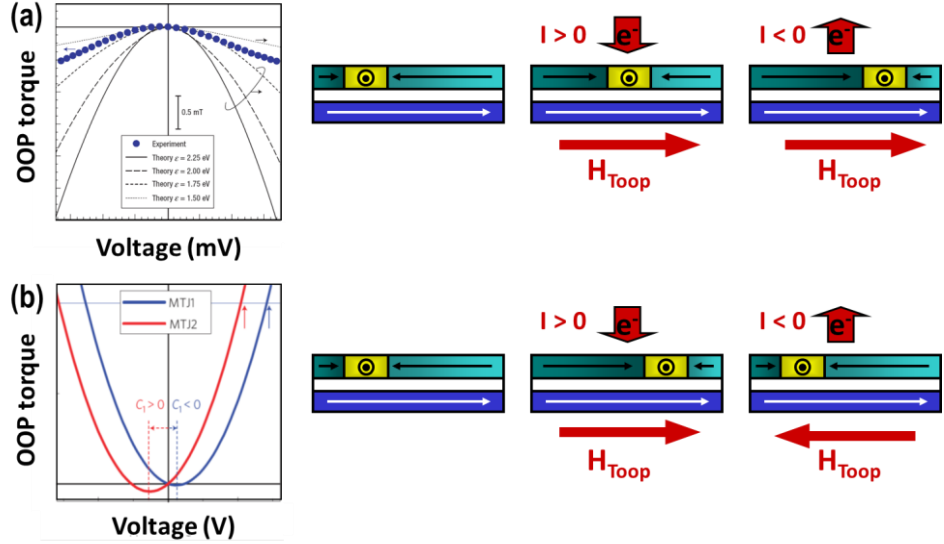

**Figure 32:** Bias dependence of the out-of-plane torque (a) in a symmetric tunnel junction (adapted from [19]) (b) in an asymmetric tunnel junction (adapted from [87]). In each case, schematic of a DW driven by the OOP torque.

In a perfectly symmetric tunnel junction, it has been shown experimentally [19] [20] and theoretically [147] that the sign of the OOP torque does not change with current (see Fig. 32a). Recent results from the Spintec group [87] supported by ab-initio calculations [148] [149], demonstrate that in asymmetric junctions with electrodes made of different materials a non-negligible linear component of the out-of-plane torque can be obtained (see Fig. 32b). For that reason, we have chosen an asymmetric stack for our spintronic memristor: [$Co_{60}Fe_{20}B_{20}$ 3 nm]$_{fixed}$ / MgO 1.1 nm / [$Co_{70}Fe_{30}$ 1 nm / $Ni_{83}Fe_{17}$ 4 nm]$_{free}$. To check the bias-dependence of $T_{OOP}$ in our junctions, we have performed "spin diode" experiments in elliptic 50 x 250 nm$^2$ tunnel junctions [88]. The results are shown on Fig. 33.

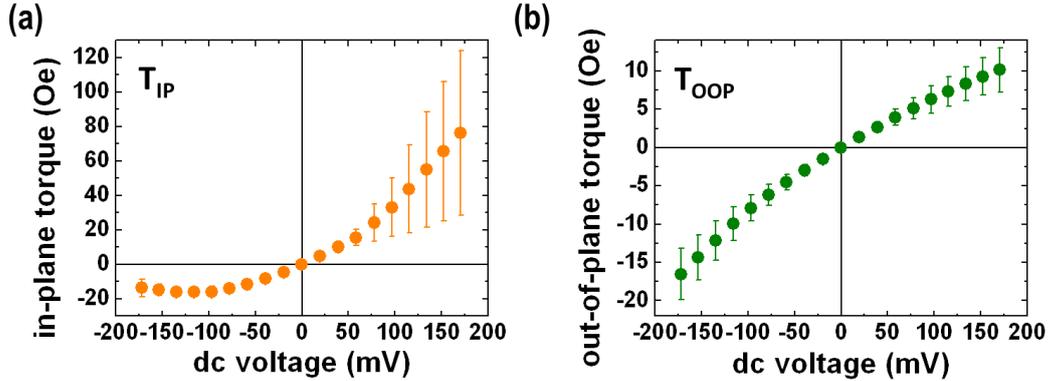

**Figure 33:** Bias dependence of the spin torques obtained from « spin diode » measurements in asymmetric CoFeB/MgO/CoFe/NiFe tunnel junctions (a) « In-Plane » torque (b) « Out-Of-Plane » torque

The in-plane torque, displayed in Fig. 33a, has the mostly linear bias-dependence expected from high TMR tunnel junctions. The OOP torque, shown on Fig. 33b, presents a surprisingly large linear component, much larger than previously measured.

We are still investigating the microscopic origin of this unconventional bias dependence. Ab-initio calculations can predict such a large linear component of $T_{OOP}$ in asymmetric systems. Different exchange splitting [148] or on-site energies of the left and right electrodes induce a linear contribution to $T_{OOP}$ [149]. However, to account for the large linear component of $T_{OOP}$ in our experiments, unrealistically large differences in the parameters characterizing the left and right electrodes have to be input (~ 1 eV difference in exchange splitting or on-site energy).

For that reason, we have searched for other possible origins of the large linear component of OOP torque in our system. We have shown theoretically with A. Manchon from KAUST that an additional mechanism can increase the linear character of $T_{OOP}$ [150]. The model takes into account spin scattering in the bulk of the electrodes. The precession and relaxation of spin accumulation induces a mixing of the torques. In particular, the out-of-plane torque steals part of the linear component of the in-plane torque. This effect is enhanced if the spin diffusion length in the electrodes is not too large compared to the dephasing ($\lambda_\perp$) and precession lengths. We are using NiFe as our electrode material, which has a small spin diffusion length (~ 5 nm) compared to materials that are traditionally used in MgO tunnel junctions such as CoFe or CoFeB (~ 60 nm). It nevertheless remains to be checked experimentally that this mechanism is at the origin of the linear enhancement of the OOP torque in our junctions.

In addition to being crucial for the spintronic memristor, we have pointed out that the linear component of $T_{OOP}$ can be used to increase of more than one order of magnitude the rectified spin diode signal (conversion form microwave power to a dc voltage), which can be very interesting for microwave-sensors [21].

**In our concept of the spin torque memristor, the driving torque for the domain wall is the out-of-plane torque $T_{OOP}$. It is therefore important to engineer a system with a large out-of-plane torque, with a linear bias dependence in order to be able to move the domain wall back and forth with currents of opposite polarity. For that purpose we have used a CoFeB / MgO / NiFe tunnel junction stack. By using the "spin torque diode" method, we have shown that the out-of-plane torque has indeed a large linear contribution in our system, even more than predicted. We have developed a model taking into account spin scattering in the electrodes to propose an origin to these experimental observations.**

**Related publications:**

R. Matsumoto, A. Chanthbouala, J. Grollier, V. Cros, A. Fert, K. Nishimura, Y. Nagamine, H. Maehara, K. Tsunekawa, A. Fukushima and S. Yuasa, "Spin-Torque Diode Measurements of MgO-Based Magnetic Tunnel Junctions with Asymmetric Electrodes", Appl. Phys. Exp., **4**, 063001 (2011)

A. Manchon, R. Matsumoto, H. Jaffres and J. Grollier, "Spin Transfer Torque with Spin Diffusion in Magnetic Tunnel Junctions", Phys. Rev. B **86**, 014402 (2012)

### d. Proof of concept

Having demonstrated that this [$Co_{60}Fe_{20}B_{20}$ 3 nm]$_{fixed}$ / MgO 1.1 nm / [$Co_{70}Fe_{30}$ 1 nm / $Ni_{83}Fe_{17}$ 4 nm]$_{free}$ stack had the required properties for our spin torque memristor : a non-zero, largely linear out-of-plane torque, we have used it to fabricate our devices. As illustrated on Fig. 34, we have processed the junctions by e-beam lithography (width ~ 200 nm, length ~ 1 µm), giving them the shape of an arc to facilitate the domain wall nucleation. As can be seen on the same figure, the vertical current injection allows displacing the DW to the left or to the right depending on the current sign. The very low critical current densities of a few $10^6$ A.cm$^{-2}$ confirm the predicted 2 orders of magnitude gain compared to the lateral injection scheme (which requires ~$10^8$ A.cm$^{-2}$ to move a DW). The currents themselves remain high, typically 10 mA, because of the large junctions area. We expect to be able to decrease them well below the mA by using a free layer with perpendicular magnetic anisotropy (domain width divided by 10 → sample area divided by more than 10).

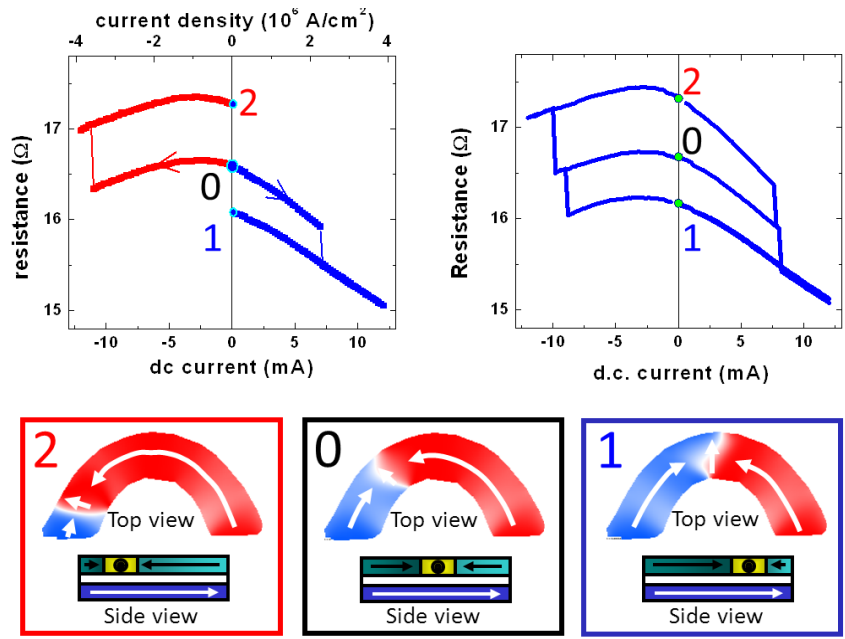

**Figure 34:** Domain wall displacement by vertical current injection. The sharp resistance variations on the resistance versus current curves correspond to the domain wall moving between the three accessible magnetic configurations labeled 0,1 and 2.

Three resistance states are obtained, corresponding to three stable positions of the DW, implementing a three state spintronic memristor. These 3 configurations, reproducible from sample to sample, are due to the combination of the specific arc shape of the junction and its terminations and the dipolar fields / coupling with the bottom layers. In the future, obtaining more states will require playing with the geometry and materials.

### e. Switching speed

A strong point of the spin torque memristor compared to other technology is its supposedly rapid switching, at the time scale of magnetization reversals, i.e. the nanosecond. In order to demonstrate this, we have performed single shot time domain measurements of the resistance variations when the domain wall is moving. The samples (see Fig. 35a) are slightly modified compared to the previous paragraph, as their shape is optimized to have only two stable domain wall positions, and the free layer is purely NiFe (5 nm).

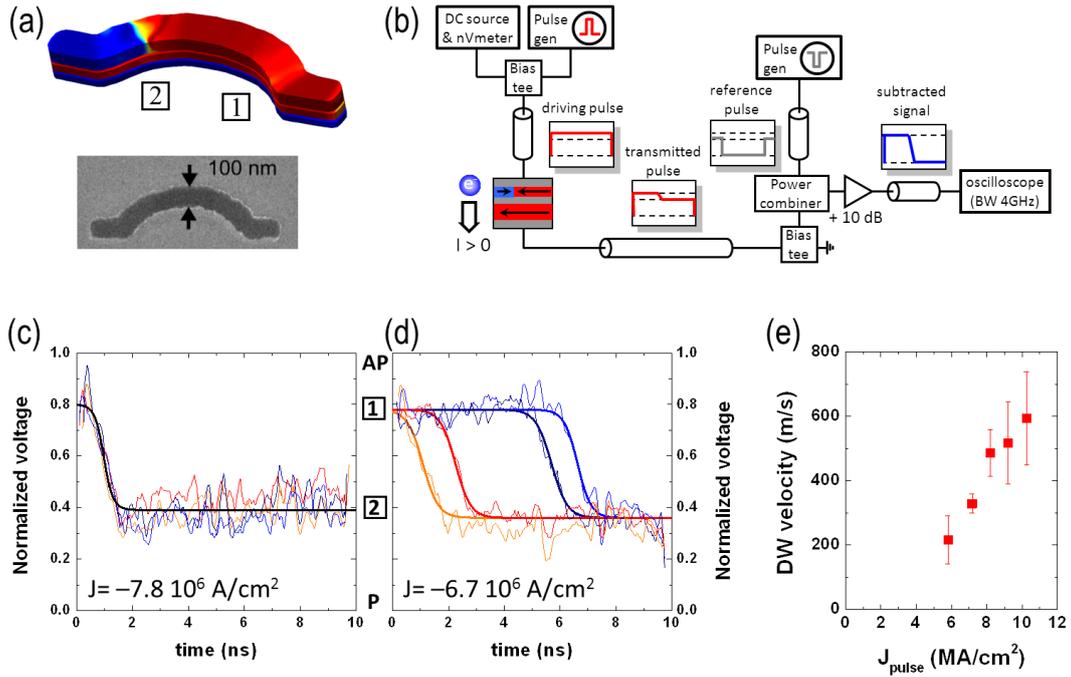

**Figure 35:** (a) Schematic of the MTJ stack showing the stable position of the DW (b) Schematic of the setup used for the time resolved measurements (c) Four single-shot voltage traces obtained under 10 ns long pulses of $J = 7.8 \cdot 10^6$ A.cm$^{-2}$. (d) Same as c, with $J = 6.7 \cdot 10^6$ A.cm$^{-2}$. (e) DW velocity as a function of the pulsed current density

A schematic of the time-resolved set-up is given in Fig. 35b. Driving pulses of 10 ns are applied to the junction. One important trick for resolving the tiny resistance variations linked to the DW motion is to subtract a reference pulse to the transmitted pulse before amplification. The single shot traces are shown in Fig. 35c and d. In each graph, four single shot traces, repeatedly obtained in the exact same conditions are displayed. In Fig. 35c, the curves, obtained for a current density of -7.8 10$^6$ A.cm$^{-2}$ are superposed. The behavior is deterministic and the DW starts moving at the onset of the pulse at t = 0. For Fig. 35d, the current density is slightly lower, of -6.7 10$^6$ A.cm$^{-2}$. The onset of domain wall motion is then scattered, providing evidence for a stochastic, thermally activated depinning. The critical current density above which deterministic domain wall motion can be achieved is therefore a few 10$^6$ A.cm$^{-2}$. In both cases the commutation is extremely fast, with switching times below the nanosecond. The domain wall velocity versus current is shown in Fig. 35e. For current densities of 10$^7$ A.cm$^{-2}$, the DW velocities exceed 600 m/s, a much higher value than can be obtained in the lateral injection geometry[2], finally approaching field-induced domain wall velocities [151].

---

[2] In the absence of spin-orbit torques.

### f. Conclusion on the spin torque memristor

> **By fabricating a 3 state spin torque memristor, we have given the proof of concept of our device. The purely electronic switching, with reasonable current densities of a few $10^6$ A.cm$^{-2}$, combined with the sub-ns commutation times, make this device very promising for applications were the number of writing cycles will be very large. To go beyond the proof of concept will require to engineer more resistance states (more DW positions), and to decrease the currents by using perpendicularly magnetized materials. The main drawback of the spin torque memristor is the low OFF/ON ratios, limited to about 6 with today's tunnel junctions, preventing their use in large crossbar arrays. Nevertheless, due to the strong research efforts to improve ST-MRAM technology, the TMR ratios keep increasing, and we are still far from the theoretical limit, which predicts $R_{AP}/R_P$ over 100 [152] [153]. In addition, the spin torque memristor has another great advantage compared to other technologies: it can be easily combined with other spin torque based devices with complementary functionalities to implement novel hybrid CMOS/Spintronics architectures. I will discuss this point in more details in the research project section.**

**Related publications:**

A. Chanthbouala, R. Matsumoto, J. Grollier, V. Cros, A. Anane, A. Fert, A. V. Khvalkovskiy, K.A. Zvezdin, K. Nishimura, Y. Nagamine, H. Maehara, K. Tsunekawa, A. Fukushima and S. Yuasa, "Vertical current induced domain wall motion in MgO-based magnetic tunnel junction with low current densities", Nature Physics **7**, 626 (2011)

J. Grollier, A. Chanthbouala, R. Matsumoto, A. Anane, V. Cros, F. Nguyen van Dau, A. Fert, "Magnetic domain wall motion by spin transfer", C. R. Physique **12**, 309–317 (2011)

## 5. Ferroelectric Memristor

### a. Solid-state ferroelectric tunnel junctions

The building block of the ferroelectric memristor is the ferroelectric tunnel junction [154]. When the screening lengths of the metallic electrodes are very different, theory predicts large resistance variations, of several orders of magnitude, when the ferroelectric polarization of the barrier commutes between up and down [155] [156] [157]. After a first set of experimental results by Contreras et al. indicating that such ferroelectric mediated resistive switching was possible [158], a definite experimental proof was brought by the "multi-functional oxide" group at UMPhy CNRs/Thales [115]. These experiments are illustrated in Fig. 36.

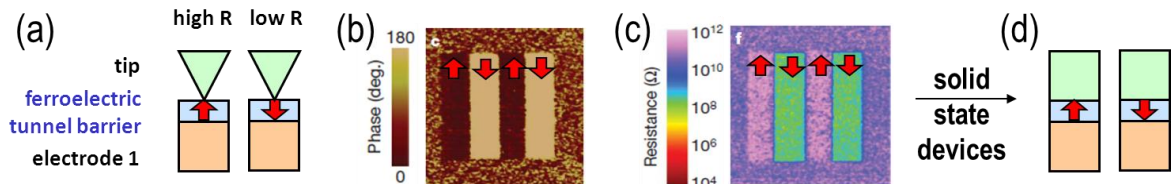

**Figure 36:** (a) Schematic of the sample geometry in Garcia et al.: the AFM tip serves as top electrode [115]. (b) phase of the piezoelectric response after having polarized the 3 nm BTO layer in alternating 1 x 4 µm$^2$ domains of up and down polarization with +/- 3.5 V writing voltage amplitude. (c) Resistance of the same sample area obtained with the conductive AFM tip. (d) Schematic of solid-state ferroelectric tunnel junctions.

The samples are highly strained ferroelectric BaTiO$_3$ (BTO) thin films (1-3 nm) deposited over 30-nm-thick La$_{0.67}$Sr$_{0.33}$MnO$_3$ (LSMO) electrodes on NdGaO$_3$ (NGO) single crystal substrates. There is no top metallic electrode on the BTO layer. Instead, a scanning probe microscope is used first to write alternating domains with up and down polarization by applying positive and negative voltages with a conductive tip AFM, then to acquire the piezoresponse of these domains (see phase map of Fig. 36b), and finally to measure the resistance of the tip/BTO/LSMO ferroelectric tunnel junction. As can be seen in Fig. 36c, the resistance of up and down domains is very different, with $R_\uparrow/R_\downarrow > 1000$. To build a ferroelectric memristor based on these giant electro-resistance effects, the first step is to fabricate solid-state ferroelectric tunnel junctions with a top electrode and demonstrate that the large resistance variations are preserved.

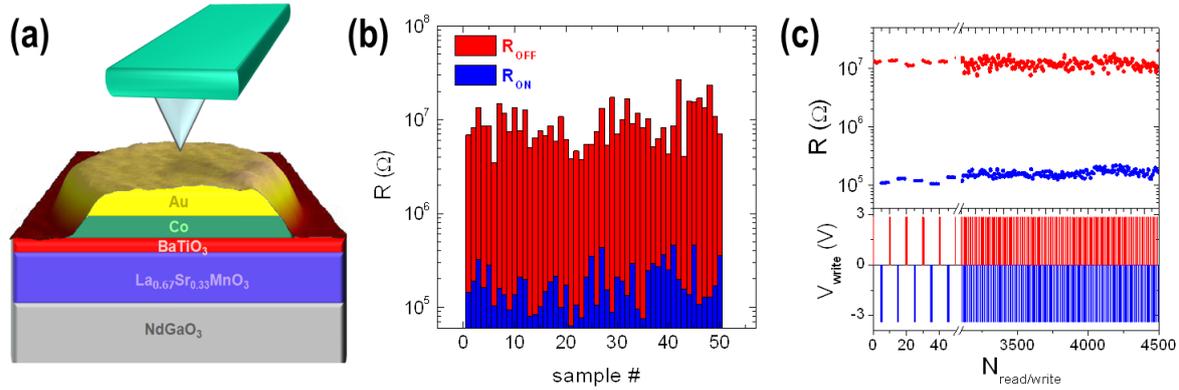

**Figure 37:** (a) Schematic of the solid-state ferroelectric tunnel junctions. (b) ON and OFF resistance states measured after applying 100 μs voltage pulses to 50 different junctions. Average OFF/ON ratio is 64 (range, 15–220) (c) Reversible resistance switching between ON and OFF resistance states of a typical junction for more than 900 cycles.

For that purpose, as illustrated on Fig. 37a, we have patterned ferroelectric tunnel junctions by depositing Co/Au dots (by e-beam lithography and subsequent lift-off) of 100 to 500 nm diameter on top of the NGO//LSMO/BTO (2 nm) stack [159]. The transport measurements are still performed by using an AFM tip, but this time for contacting the top Au electrode. Voltage pulses of a few volts (typically below 5 V) and variable duration (10 ns to 100 μs) are applied across the junction. The resistance state is then read at remanence by applying a small sub-threshold voltage (~ 100 mV). As shown in Fig. 37b, the OFF and ON states are fairly reproducible from sample to sample. The $R_{OFF}/R_{ON}$ ratio, about 100, is a bit lower than measured with the AFM tip, probably due to very different screening lengths of the diamond tip coating compared to cobalt. We have also performed a first measurement of the endurance of our devices, limited due to the drift of the AFM tip. Nevertheless, we can say that the two resistance states stay stable over 900 writing cycles (see Fig. 37c). In addition, we have shown that our devices can be switched with 10 ns pulses, which are the shorter pulses we can transmit through the AFM tip and its cables.

### c. Ferroelectric memristor: proof of concept

The samples are therefore a good starting base for implementing a ferroelectric memristor. To obtain the quasi-analog resistance variations we are again going to exploit the domain configuration, this time of the ferroelectric barrier [160], as illustrated in Fig. 38.

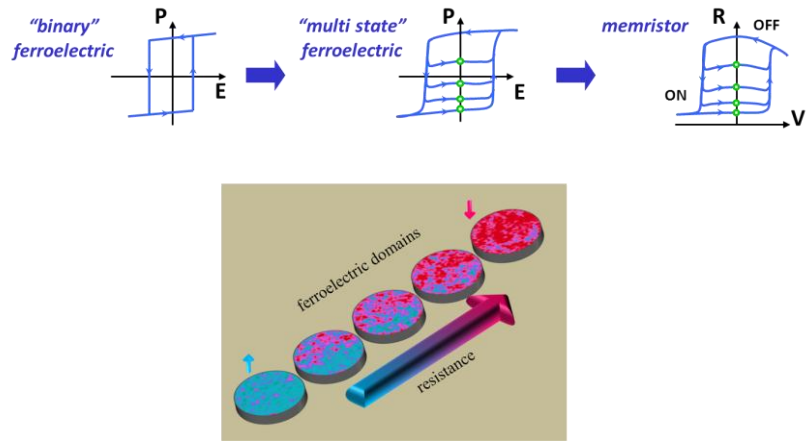

**Figure 38:** Principle of the ferroelectric memristor. The multi-state quasi-analog resistance variations will be obtained thanks to the nucleation and propagation of ferroelectric domains in the barrier.

For the spin torque memristor, we were using in-plane magnetized ferromagnetic layers, for which the domain size can hardly be shrunk below 100 nm. For that reason, we chose to modulate the resistance via a single domain wall, which requires a very precise control of its position through spin torque induced dynamics. The advantage of ultra-thin ferroelectric layers is the possibility to stabilize very small ferroelectric domains. Indeed, while the domain size of ferroic materials scales as the square root of the film thickness [161], the proportionality constant, which depends on the domain wall width, is much smaller for ferroelectric than for ferromagnetic materials [162] [163]. In our 2 nm BTO barrier, the expected domain size is below 5 nm [164]. Therefore, in our dots of lateral dimensions around 100 nm, the polarization switching will not be sharp, but will occur progressively through the electric-field dependent nucleation of domains, and subsequent propagation of domain walls. This will allow, as illustrated in Fig. 38, to obtain a voltage tunable, quasi-analog variation of device resistance.

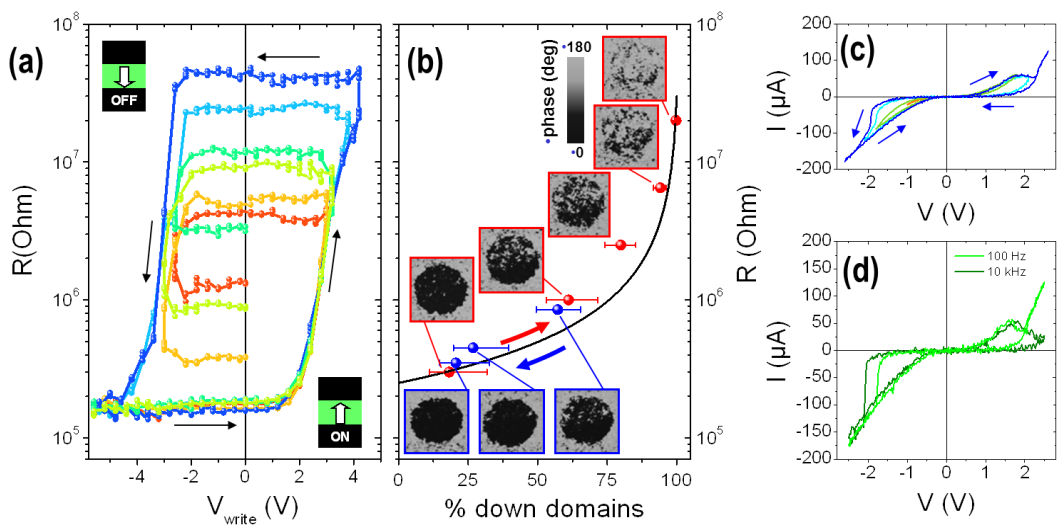

**Figure 39:** (a) Dependence of the junction resistance after the application of 20 ns voltage pulses ($V_{write}$) of different amplitudes. (b) Variation of a similar capacitor resistance with the relative fraction of down domains extracted from the PFM phase images. Red-(and blue-) framed images show states achieved by the application of positive (and negative) voltage pulses of increasing amplitude starting from the ON (and OFF) state. The blue and red symbols correspond to the experimental resistance value as a function of the fraction of down domains extracted from the PFM phase images; the black curve is a simulation in a parallel resistance model. (c,d) Current versus voltage curves, measured at 1 kHz on a similar capacitor, for various amplitudes of the maximum voltage (c) and current versus voltage curves, measured at different frequencies: 100 Hz and 10 kHz (d)

In order to check experimentally this idea, we have measured the resistance versus voltage characteristics of our junctions. We apply voltages pulses of 20 ns duration and varying amplitude, and read the resistance at remanence ($V_{read}$ = 100 mV) after each pulse. As shown in Fig. 39a a hysteretic cycle between low ($R_{ON}$ ~ 1.6 $10^5$ Ω) and high ($R_{OFF}$ ~ 4.6 $10^7$ Ω) resistance states is observed, with a large OFF/ON ratio of 300. The switching between the two states is bipolar and, as expected, not abrupt: a broad range of intermediate resistance states are observed. The minor loops (cyan to red curves) show that depending on the cycling protocol the final resistance state can be finely tuned between $R_{ON}$ and $R_{OFF}$: here the hysteresis of ferroelectric switching confers the memristor its memory effect.

An advantage of our experimental configuration is that we can image the details of the ferroelectric domain structure through the Co/Au pad by piezo-force response microscopy (PFM). The resulting images, acquired at different voltage values are shown in Fig. 39b. Starting from a uniform state, the application of voltage pulses with increasing amplitude nucleates then propagates domains of opposite polarity. The resistance of the junction is measured after each image. In that way, it is possible to plot the measured resistance as a function of the proportion of up and down domains extracted from each image. As shown in Fig. 39b, the junction resistance shows a systematic variation with the relative fraction of down domains extracted from the PFM images, well reproduced by a simple model of parallel conduction for the up and down domains (line in Fig. 39b). The memristive character of the junctions is further confirmed by the current versus voltage curves presented in Fig. 39c,d. The observed IV loops are pinched, as expected for a memristor device. They expand as the maximum voltage increases (Fig. 39c), and as the frequency is increased (Fig. 39d).

> **The ferroelectric memristor is based on the gradual switching of polarization in the ferroelectric barrier of a solid-state tunnel junction. When the thickness of a ferroelectric layer is decreased below a few nanometers, the domain size becomes very small as well, typically a few nm. We have demonstrated by imaging the ferroelectric domain configuration during switching that the quasi-analog resistance variations in our memristor are obtained through the formation and expansion of ferroelectric domains. The large OFF/ON ratios over two orders of magnitude, fast switching below 10 ns, combined to the purely electronic operation, make the ferroelectric memristor an excellent candidate for future integration in large scale crossbar arrays.**

**Related publications:**

A. Chanthbouala, V. Garcia, R. O. Cherifi, K. Bouzehouane, S. Fusil, X. Moya, S. Xavier, H. Yamada, C. Deranlot, N. D. Mathur, M. Bibes, A. Barthélémy and J. Grollier, "A ferroelectric memristor", Nature Materials **11**, 860-864 (2012)

A. Chanthbouala, A. Crassous, V. Garcia, K. Bouzehouane, S. Fusil, X. Moya, J. Allibe, B. Dlubak, J. Grollier, S. Xavier, C. Deranlot, A. Moshar, R. Proksch, N. D. Mathur, M. Bibes and A. Barthélémy, "Solid-state memories based on ferroelectric tunnel junctions", Nature Nano. **7**, 101 (2012)

### d. Switching dynamics

The above experimental results confirm that, in ferroelectric tunnel junctions, the memristive response can be devised by controlling the nucleation and growth of ferroelectric domains. In order to get a better insight in the dynamics at stake, we have studied the evolution of the fraction of reversed domains, obtained from the resistance value, as a function of the cumulated pulse time. In our samples, the low-resistance state ($R_{ON}$) corresponds to the ferroelectric polarization pointing up ($P_\uparrow$), that is, towards the Co/Au pad, which is also the virgin state for all devices. The relative fraction of down domains can be defined by s = (1/R-1/$R_{ON}$)/(1/$R_{OFF}$-1/$R_{ON}$); thus, s varies from 0 in the ON state ($P_\uparrow$) to 1 in the OFF state ($P_\downarrow$). Fig. 40 shows a typical set of data on the evolution of s as a function of cumulative pulse time for pulse durations of 10 ns.

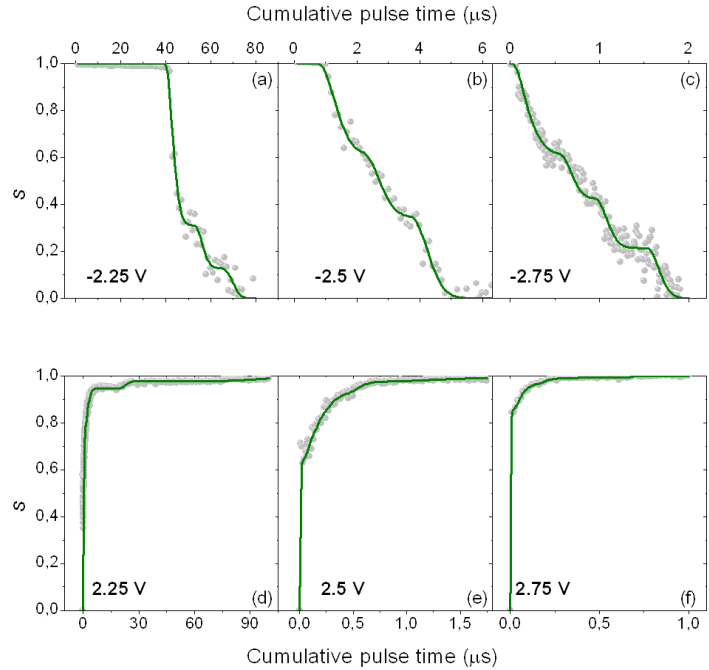

**Figure 40:** Polarization switching dynamics. (a–f): Dependence of the switched fraction on the cumulative pulse time for down-to-up (a–c) and up-to-down (d–f) switching and different voltage amplitudes. The data are shown as symbols and the lines are fits.

For positive (negative) pulses, the initial state was initialized to $R_{ON}$ ($R_{OFF}$) corresponding to $P_\uparrow$ ($P_\downarrow$). While the polarization reversal starts immediately after the first pulse for up-to-down switching (Fig. 40d-f), it is delayed in the down-to-up case with a delay time that depends on the applied voltage (Fig. 40a-c). This behavior is consistent with the visible shift in the R(V) cycle of Fig. 39a, that may reflect the presence of downward-polarized interfacial dipoles favouring the initial growth of ↓ domains [165]. The delayed onset of switching at negative voltage can then be ascribed to asymmetric nucleation processes: for up-to-down switching, pinned domains with down polarization serve as pre-existing nucleation centers; on the contrary, for down-to-up switching, nucleation centers need to be activated, explaining the observed delays in the s vs. time data (Figs. 40a-c) corresponding to increased nucleation times. For both switching directions, s does not always evolve smoothly toward the final state but presents a more "wavy" dependence. This signals the presence of several areas with different switching dynamics. These spatial inhomogeneities could be due to the sub-micron lithographic process we use to define FTJs that may introduce a slight polarization disorder.

In order to account for these dynamics, we have slightly modified the Kolmogorov-Avrami-Ishibashi (KAI) model [166] of ferroelectric reversal via domain nucleation and subsequent DW propagation. We have modeled the data by dividing the pad area in a finite number of zones with different propagation and nucleation kinetics (different domain wall propagation speeds, nucleation times, number of nuclei), each ruled by the KAI model. Figure 40 also shows the fit of the experimental data by our extended KAI model. The data are well fitted on the whole time range; in particular the "wavy" dependence of s vs. time is accurately reproduced with a reduced number of zones N ≤ 5.

This good agreement between model and experimental data is particularly important. Indeed, it will also us to describe and predict the memristive response with physical equations. This is not the case with most resistive switching memristors, where the microscopic phenomena behind the resistance variations are poorly understood, and extremely complex, only allowing phenomenological descriptions of final device behaviors.

> **We have extended the KAI model of domain nucleation and subsequent domain wall propagation to account for the experimentally observed ferroelectric polarization dynamics. This is an important step to engineer and predict the memristive response.**

### e. Conclusion on the ferroelectric memristor

We have developed a ferroelectric memristor with purely electronic commutation and large resistance variations at the time scale of tens of nanoseconds. The perspectives are double. From the device point of view, we will have to fabricate all-solid state pillar junctions with a top electrode. This is an important step before the fabrication of ferroelectric memristive crossbars. These junctions will be measurable with classical transport characterization method, without the need of the AFM set-up. This will open the path to sub-ns dynamical measurements, endurance tests etc. From the physics point of view, we now have a tool to probe by simple transport measurements the fast dynamics of ferroelectric domain walls. It will be interesting to investigate these dynamics in other materials, and to study the impact of junction geometry on the dynamic response.

## 6. Conclusion on memristors

We have given the proof of concept of our two purely electronic memristors: the spin torque and ferroelectric memristor. For each type of device, the next step is to demonstrate their potentiality as artificial synapse by integrating them in a small hardware neural network demonstrator.

# D. RESEARCH PROJECT: MULTI-FUNCTIONAL NANO-DEVICES FOR BIO-INSPIRED COMPUTING

# E. BIBLIOGRAPHY

*Jpn.,* vol. 31, p. 506, 1970.